\documentclass[usenatbib]{mnras}
\usepackage{amsmath}
\usepackage{amssymb}
\usepackage{graphicx}
\usepackage[utf8]{inputenc}
\usepackage{listings}
\usepackage{geometry}
\usepackage[normalem]{ulem}
\usepackage{xcolor}
\usepackage{booktabs,tabularx}
\usepackage{tablefootnote}
\usepackage{hyperref}
\usepackage[toc,page]{appendix}
\hypersetup{colorlinks=true,linkcolor=blue,citecolor=blue,filecolor=blue,urlcolor=blue}

\def \be{\begin{equation}}
\def \ee{\end{equation}}
\def \bea{\begin{eqnarray}}
\def \eea{\end{eqnarray}}

\def\kms{\ifmmode {\rm km\:s^{-1}} \else $\rm km\:s^{-1}$\fi}
\def\mbar{\ifmmode {\bar{m}} \else $\bar{m}$\fi}
\def\vturb{\ifmmode {\sigma_{\rm turb}} \else $\sigma_{\rm turb}$\fi}
\def\kb{\ifmmode {k_{\rm B}} \else $k_{\rm B}$\fi}
\def\mp{\ifmmode {m_{\rm p}} \else $m_{\rm p}$\fi}
\def\rcgm{\ifmmode {r_{\rm CGM}} \else $r_{\rm CGM}$\fi}

\defcitealias{Faerman2017}{FSM17}
\defcitealias{faerman2020}{FSM20}

\begin{document} 
\title[CGM model comparison project]{Comparison of Models for the Warm-Hot Circumgalactic Medium around Milky Way-like Galaxies}

\author [P. Singh et al.]
{Priyanka Singh$^{1,2,3}$ \thanks{psingh@iiti.ac.in}, 
Erwin T. Lau$^{4,5}$  \thanks{erwin.lau@cfa.harvard.edu}, 
Yakov Faerman$^{6}$,
Jonathan Stern$^{7}$,
Daisuke Nagai$^{1,2}$\\
$^{1}$Yale Center for Astronomy \& Astrophysics, New Haven, CT 06511, USA \\
$^{2}$Department of Physics, Yale University, New Haven, CT 06520, USA \\
$^{3}$Department of Astronomy, Astrophysics and Space Engineering, Indian Institute of Technology, Indore 453552, India \\
$^{4}$Center for Astrophysics $\vert$ Harvard \& Smithsonian, 60 Garden St, Cambridge, MA 02138, USA\\
$^{5}$Department of Physics, University of Miami, Coral Gables, FL 33124, USA\\
$^{6}$Astronomy Department, University of Washington, Seattle, WA 98195, USA\\
$^{7}$School of Physics \& Astronomy, Tel Aviv University, Tel Aviv 69978, Israel
}

\label{firstpage}
\maketitle

\begin{abstract}
A systematic comparison of the models of the circumgalactic medium (CGM) and their observables is crucial to understanding the predictive power of the models and constraining physical processes that affect the thermodynamics of CGM. This paper compares four analytic CGM models: precipitation, isentropic, cooling flow, and baryon pasting models for the hot, volume-filling CGM phase, all assuming hydrostatic or quasi-hydrostatic equilibrium. We show that for fiducial parameters of the CGM of a Milky-Way (MW) like galaxy ($\rm M_{vir} \sim 10^{12}~M_{\odot}$ at $z\sim 0$), the thermodynamic profiles -- entropy, density, temperature, and pressure -- show most significant differences between different models at small ($r\lesssim 30$ kpc) and large scales ($r\gtrsim 100$ kpc) while converging at intermediate scales. The slope of the entropy profile, which is one of the most important differentiators between models, is $\approx 0.8$ for the precipitation and cooling flow models, while it is $\approx0.6$ and 0 for the baryon pasting and isentropic models, respectively. We make predictions for various observational quantities for an MW mass halo for the different models, including the projected Sunyaev-Zeldovich (SZ) effect, soft X-ray emission (0.5--2 keV), dispersion measure, and column densities of oxygen ions (OVI, OVII, and OVIII) observable in absorption. 
We provide Python packages to compute the thermodynamic and observable quantities for the different CGM models.

\end{abstract}


\section{Introduction}
\label{sec-intro}

A large fraction of baryons associated with galactic halos reside in a gaseous phase, extending out and potentially beyond the virial radius of the halo. This gaseous halo is referred to as the intra-cluster medium (ICM) in clusters of galaxies, the intra-group medium (IGrM) in galaxy groups, and the circumgalactic medium (CGM) around galaxies \citep[see][for review]{Tumlinson2017}. Amongst these, the CGM is the most poorly constrained regime observationally owing to its lower density and temperature than the IGrM and ICM and theoretically due to the major impact of non-gravitational processes such as feedback and turbulence. 

The CGM can be studied across multiple wavelengths, ranging from microwave \citep{PZ19}, UV \citep{Werk2014, Lehner2015, Qu2018, Chen20, Tchernyshyov2022}, to X-rays \citep{Anderson2016, Li2018, Das2021}. More recently, \cite{Bregman2022} detected resolved thermal Sunyaev-Zeldovich (tSZ) profiles from $L^*$ galaxies, constraining their hot baryon budget. \cite{Chadayammuri2022} and \cite{Comparat2022} stacked star-forming and passive galaxies in the eROSITA Final Equatorial Depth Survey (eFEDS), and \cite{Zhang2024} stacked central and isolated galaxies in the first data release of the eROSITA all sky survey (eRASS), measuring the resolved X-ray surface brightness profiles from the CGM. These observations have pushed the detection capabilities to new limits (i.e., resolving the radial distribution of the CGM down to Milky-Way (MW) masses). 

Both the quality and quantity of CGM measurements are expected to take another leap with ongoing and upcoming experiments, such as the Canadian Hydrogen Intensity Mapping Experiment (CHIME, \citealt{Amiri2018}) and the Hydrogen Intensity and Real-time Analysis eXperiment (HIRAX, \citealt{Newburgh2016}) in the radio, detecting the dispersion measure from Fast Radio Bursts (FRBs), the Advanced Atacama Cosmology Telescope (AdvACT, \citealt{Henderson2016}), South Pole Telescope-3G (SPT-3G, \citealt{Benson2014}), Simons Observatory \citep{Ade2019}, and CMB-S4 \citep{Abazajian2016} at mm wavelength detecting the SZ effect, and eROSITA All Sky Survey (eRASS) in the X-ray. Through SZ and X-ray surveys, we will probe the resolved CGM profiles to the virial radii of $M_{\rm vir} \sim 10^{12}~{\rm M}_{\odot}$ galaxies. At the same time, ongoing and upcoming FRB observations will push the mass limit down to $M_{\rm vir} \sim 10^{11}~{\rm M}_{\odot}$ \citep{Battaglia2019, Wu2022}. Therefore, on the theoretical modelling front, we need to prepare ourselves to maximise the CGM physics extracted through these next-generation missions.

Several theoretically and observationally motivated models have been developed to describe and study the dominant physical processes that govern the CGM \citep[e.g.,][]{Voit2017, Choudhury2019, Stern2019, faerman2020, singh2021, Pandya2022}. These models represent a simplified approach to modelling CGM thermodynamics. Hydrodynamical cosmological simulations, on the other hand, capture a more realistic and complex interplay between the different processes in the CGM \citep[e.g.,][]{Oppenheimer2018, Hafen2019, Hummels2019, Peeples19, Voort19, Ramesh2023}. Several studies have made comparisons of hydrodynamical simulations from the CGM scale to the ICM scale \citep{Lim2021, Lee2022, Yang2022}. More recently, using CAMELS simulations with varying feedback parameters in a variety of subgrid physics modules \citep{Villaescusa2021, Ni2023, Lee2024}, it has become possible to systematically explore the impact of feedback physics on CGM observables, such as the tSZ effect \citep{Moser2022}, X-ray \citep{Contreras2023}, and FRB \citep{Medlock2024}. 

Understanding how feedback impacts the CGM observables is complex because of the interplay of physical processes in hydrodynamical simulations. Exploring the parameter spaces of feedback physics using these simulations is also computationally expensive. Idealised analytical CGM models, on the other hand, can efficiently isolate the impact of specific physical processes. Therefore, both idealised models and hydrodynamical simulations are crucial for accurately modelling gas physics and improving our understanding of CGM and its role in galaxy evolution. Comparison of different CGM models in the literature can be challenging because they are based on different input assumptions, such as the underlying DM halo potential, models of gas cooling, and metal distributions.

In this study, we compare four idealised Milky Way-like CGM models that represent different key aspects of CGM physics. The goal is to determine whether upcoming multi-wavelength observations can differentiate between these models. The models being compared are the precipitation model \citep{voit2018_precipitation, voit2019_column_densities, singh2021}, isentropic model \citep{faerman2020,Faerman2022}, cooling flow model \citep{Stern2019,Stern2020, Stern2023}, and the baryon pasting model \citep{Shaw2010,Flender2017,osato22}. To facilitate an efficient and meaningful comparison between these different CGM models, we have developed a standardised Python pipeline to input the models and compute observables consistently. Our main objectives are to: i) compare different CGM models in a standardised manner, ii) highlight inherent differences arising from the different implementations of physical processes governing CGM physics, and iii) provide the scientific community with a user-friendly pipeline that can be expanded to include additional models.

The organisation of the paper is as follows. Section~\ref{sec-analytical-cgm} briefly introduces the conservation equations of mass, momentum, and energy that govern CGM thermodynamics and the four idealised CGM models that we address. In Section~\ref{sec-thermodynamic-profiles}, we compare the entropy, pressure, density, and temperature profiles for the fiducial parameter values of these CGM models. Section \ref{sec-multi-wavelength} describes the observational predictions, such as SZ, X-ray surface brightness, oxygen column densities, and dispersion measure profiles. We summarise the results of our analysis in Section~\ref{sec-summary}.

\section{Idealized one-dimensional models of the CGM}
\label{sec-analytical-cgm}

\subsection{General Framework}
For the hot, diffuse CGM, we can reasonably assume that the gas is collisional with low viscosity, given that the mean free path is small for the weakly magnetised plasma. The thermodynamical properties of a collisional inviscid fluid in a gravitational potential can generally be described by the three equations that represent the conservation of mass, momentum, and energy (or entropy). 

The one-dimensional radial equations for mass and momentum conservation for the inviscid CGM are,
\begin{equation}
     \frac{\partial \rho}{\partial t}+\frac{\partial}{\partial r}\left(\rho \, v_r\right)=0,
\label{eqn-mass-conservation}
\end{equation}

\begin{equation}
    \frac{\partial v_r}{\partial t} + v_r \frac{\partial v_r}{\partial r} = -\sum_i \frac{1}{\rho}\frac{\partial P_i}{\partial r} - \frac{\partial \Phi}{\partial r},
\label{eqn-momentum-conservation}
\end{equation}
where $\rho$ is gas density, $v_r$ is the radial velocity component, $P_i$ is $i^{th}$ gas pressure component, and $\Phi$ is the gravitational potential. The specific angular momentum is assumed to be zero for the CGM models considered in this paper. The total gas pressure can be written as the sum of contributions from thermal pressure, turbulence, magnetic fields, and cosmic rays (i.e., $P = \sum_i P_i = P_{th} + P_{turb} + P_{B} + P_{CR}$), where different pressure components correspond to different polytropic indices, $\gamma_i$.

The one-dimensional energy conservation equation, again assuming negligible thermal conduction and viscosity, is expressed as
\begin{eqnarray}
\frac{\partial}{\partial t}\left[\rho\left(e + \frac{v_r^2}{2}\right)\right] + \frac{\partial}{\partial r}\left[\rho\left(e + \frac{v_r^2}{2}\right)v_r + P v_r\right] \\ \nonumber = -\rho v_r\frac{\partial \Phi}{\partial r} + \mathcal{H}-\mathcal{C},
\label{eqn-energy-conservation-1}
\end{eqnarray}
where $e$ is the velocity dispersion, $\mathcal{H}$ and $\mathcal{C}$ represent the non-adiabatic heating and cooling 
per unit volume, respectively: heating occurs at the accretion shock at the outer boundary of the halo, at merger shocks, as well as through feedback at the halo core and through turbulent dissipation throughout the volume of the halo. Cooling is mainly driven by metallicity-dependent radiative cooling. 
Alternatively, Equation 3 can be rewritten as the conservation of entropy $K \equiv P \rho^{-\gamma}$:
\begin{equation}
    \left(\frac{P}{\gamma-1}\right)\frac{\partial \ln K}{\partial t} + v_r\left(\frac{P}{\gamma-1}\right)\frac{\partial \ln K}{\partial r} = \mathcal{H}-\mathcal{C},
\label{eqn-energy-conservation-2}
\end{equation}
The equations described above govern the thermodynamics and kinematics of the CGM, and can be reduced to simpler forms under certain assumptions about the CGM properties. We now discuss these for different models.

\subsection{Precipitation model}
The precipitation-limited hot halo model (simply precipitation model henceforth) \citep{voit2018_precipitation, voit2019_column_densities, singh2021} assumes the halo gas is in hydrostatic equilibrium, with no large-scale ordered inflows or outflows, i.e., $v_r \approx 0$ in Equations~\ref{eqn-mass-conservation} and \ref{eqn-momentum-conservation}, setting the components on the left-hand side of Equation~\ref{eqn-momentum-conservation} to zero.
The key ingredient of the precipitation model is a fixed ratio of gas cooling to free-fall timescales ($t_{\rm cool}/t_{\rm ff}$) throughout the halo, where $t_{\rm ff}=\sqrt{2} \times r/v_{\rm c}$ is the free-fall time, $v_{\rm c}=\sqrt{GM(<r)/r}$ is the circular velocity, and $t_{\rm cool}\equiv(\gamma-1)^{-1}P_{\rm th}/\mathcal{C}$ is the cooling time.  

The entropy profile of the precipitation model is a sum of two components, a baseline entropy profile ($K_{\rm base}$) and a precipitation limited entropy profile ($K_{\rm pre}$) given by

\begin{equation}
\label{eqn-kbaseline}
  K_{\rm base}(r) = 1.32 \, \frac{k T_{\phi}(R_{200})}{\bar{n}^{2/3}_{e,200}} 
                        \left( \frac{r}{R_{200}}\right)^{1.1} \; \; ,
\end{equation}
\begin{equation}
\label{eqn-kpre}
  K_{\rm pre}(r) = (2 \mu m_p)^{1/3} \left[\Bigl(\frac{t_{\rm cool}}{t_{\rm ff}}\Bigr) \frac{2n_i}{n}
        \frac {\Lambda(2T_{\phi},Z)} {3} \right]^{2/3} r^{2/3}
        \; \; .
\end{equation}
Here, $T_{\phi}$ is the gravitational temperature of the halo ($k T_{\phi} \equiv \mu m_p v^2_c(r)/2$), $\bar{n}_{e,\rm 200}$ is the mean electron density corresponding to 200 times the critical density and $\Lambda(T,Z)=\mathcal{C}/n^2_{\rm H}$ is the cooling function which depends on gas temperature and metallicity. 

\begin{figure*}
\centering
 \includegraphics[height=6cm,angle=0.0 ]{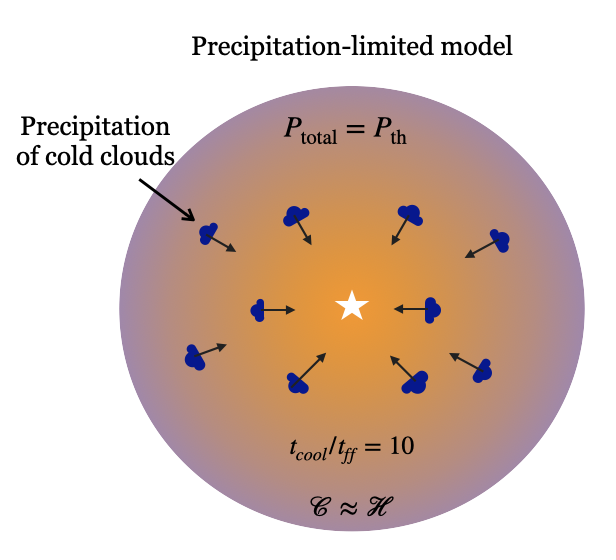}
 \includegraphics[height=6cm,angle=0.0 ]{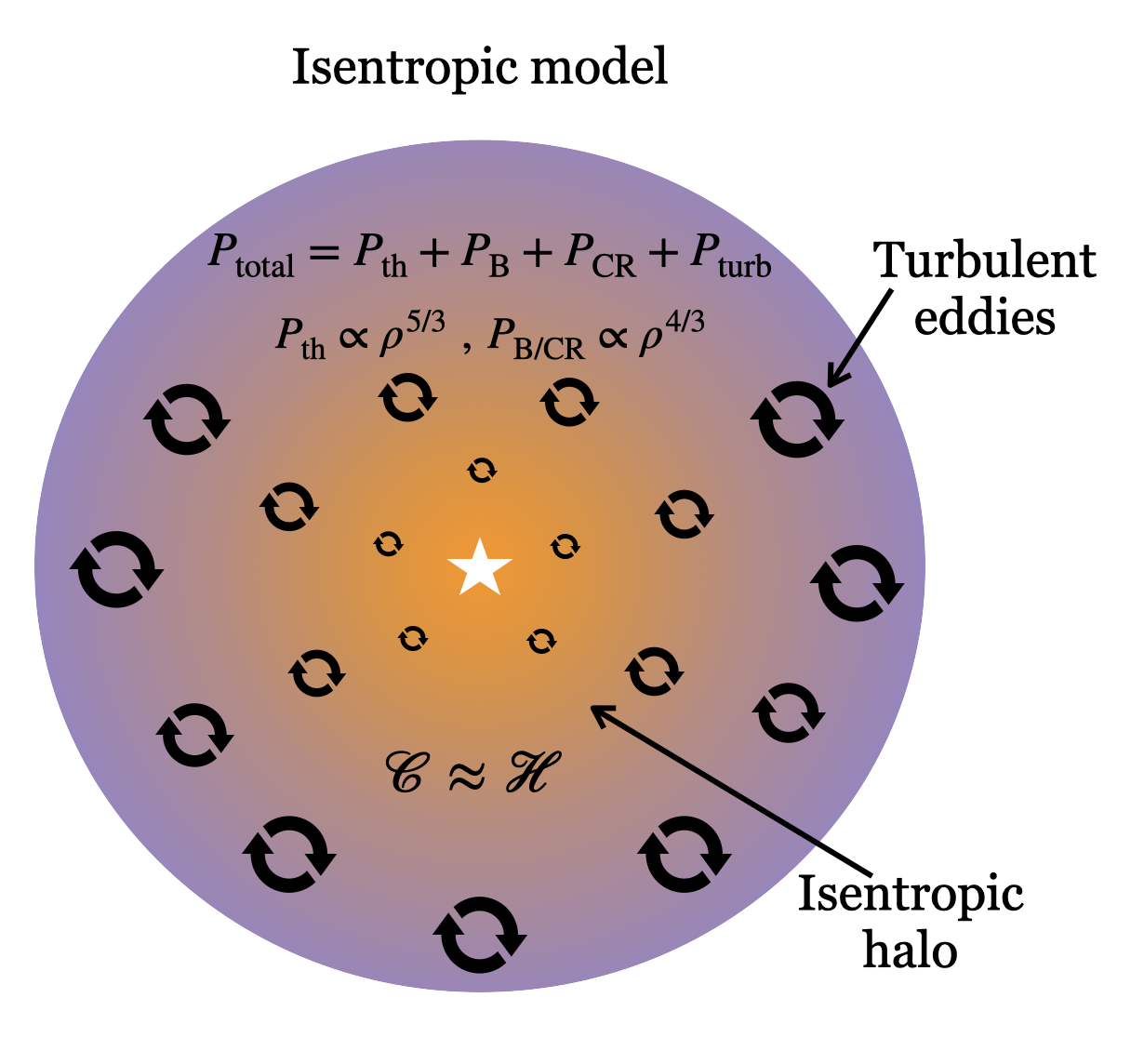}
 \includegraphics[height=6.2cm,angle=0.0 ]{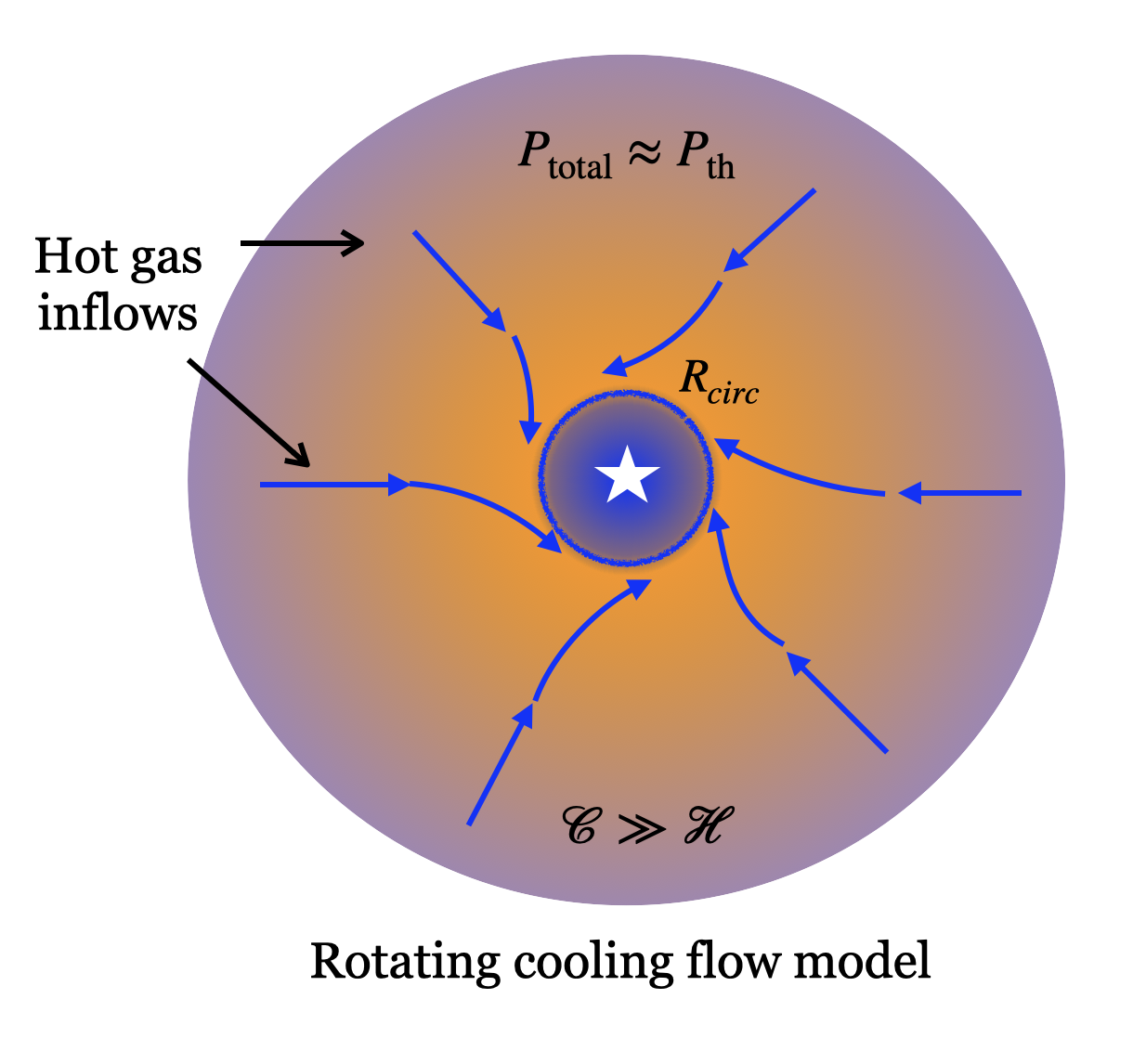}
 \includegraphics[height=6cm,angle=0.0 ]{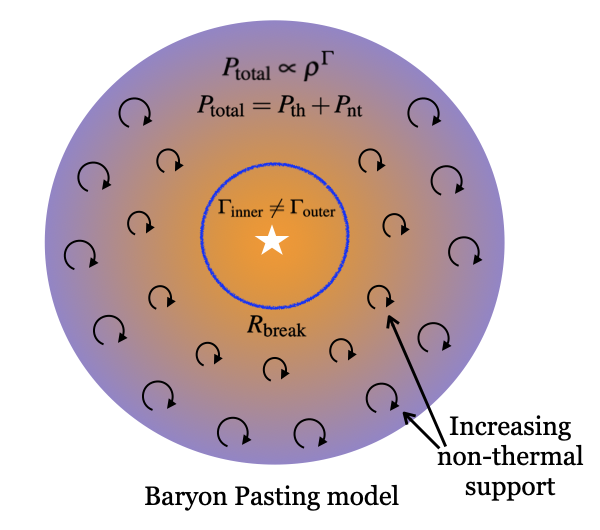}
	\caption{An illustration of the idealized CGM models considered in this work. Orange indicates a higher gas temperature than blue. {\bf Top-left (precipitation model):} thermal balance ($\mathcal{C} \approx \mathcal{H} $) is maintained while the ratio $t_{\rm cool}/t_{\rm ff}$ is fixed above a critical value of 10 throughout the halo. In cases where $t_{\rm cool}/t_{\rm ff}<10$, precipitation of cold clouds onto the central galaxy fuelling star formation and central supermassive black hole takes place. This is followed by feedback processes, which regulate the system back to thermal balance. {\bf Top-right (isentropic model):} The thermal balance is maintained throughout the halo. The nonthermal contribution to the total pressure increases with increasing galactocentric radii. Among the non-thermal components, the relative fraction of turbulent support increases faster. {\bf Bottom-left (cooling flow model):} Cooling dominates non-adiabatic heating throughout the halo. A hot inflow develops down to the circularization radius $\rm R_{circ}$, at which the hot inflow cools and fuels star formation. {\bf Bottom-right (baryon pasting model):} The relation between CGM pressure and density is controlled by a polytropic index $\Gamma$. The impact of the cooling core is captured by a break in the value of $\Gamma$ at $R_{\rm break}$, where $\Gamma_{\rm inner} \ll \Gamma_{\rm outer}$. The nonthermal pressure increases with increasing galactocentric radii.}
	\label{fig-illustration}
\end{figure*}

The baseline entropy profile in Equation~\ref{eqn-kbaseline} is a fit to simulated clusters from gravity-only cosmological simulations in the radial range $\rm (0.2-1) \times R_{200}$ \citep{Voit2005}. It represents the non-adiabatic gravitational heating from accretion shocks $\mathcal{H}$ in Equation~\ref{eqn-energy-conservation-2}. 
The value of $t_{\rm cool}/t_{\rm ff}$ controls the precipitation-limited entropy profile in Equation~\ref{eqn-kpre}, constraining the metallicity-dependent gas cooling $\mathcal{C}$ in Equation~\ref{eqn-energy-conservation-2}. The precipitation model also considers only the contribution to the gas pressure of the thermal component. Figure \ref{fig-illustration} (top left) illustrates a thermal balance when condition $t_{\rm cool}/t_{\rm ff}\gtrsim 10$ is satisfied. The precipitation of cold clouds onto the central galaxy follows as $t_{\rm cool}/t_{\rm ff}$ falls below this critical value. Precipitation, in turn, fuels next-generation feedback processes, thus restoring the system to thermal balance.

In summary, the precipitation model attempts to portray a picture of a gaseous halo in hydrostatic equilibrium with its host halo, while gas hydrodynamics on global scales is mainly governed by the ratio of $t_{\rm cool}$ and $t_{\rm ff}$. It provides an upper limit on the gas density, which then translates to the upper limit on the observed X-ray luminosity temperature relation \citep{voit2018_precipitation} from individual galaxies to massive galaxy clusters, i.e., three orders of magnitudes in halo masses ($\sim 10^{12}-10^{15} M_{\odot}$). The observed precipitation limit corresponds to a lower limit on $t_{\rm cool}/t_{\rm ff} \sim 10$.

\subsection{Isentropic model}

The isentropic model, presented in \citet[hereafter FSM20]{faerman2020}, describes a large-scale, spherically symmetric corona, with gas in hydrostatic equilibrium in the gravitational potential of a MW mass dark matter halo. The model is motivated by galactic feedback heating,  from AGN or star formation, leading the CGM to evolve toward marginal convective equilibrium. The model, therefore, adopts an adiabatic equation of state (EoS), $P = K \rho^{\gamma}$, where $K$ is the entropy parameter, constant with radius. The model allows for three pressure components: (i) thermal, (ii) nonthermal from magnetic fields and cosmic rays (B/CR), and (iii) turbulent support. Polytropic indices are $\gamma_1 = 5/3$ for the thermal pressure and $\gamma_2 = 4/3$ for the B/CR component, modelled as a relativistic fluid. The model assumes a constant velocity scale for the turbulent component, $\sigma_{\rm turb}$, corresponding to $\gamma_3 = 1$.

The model assumes that there are no large-scale ordered inflows or outflows ($v_r \approx 0$ in Equations~\ref{eqn-mass-conservation} and \ref{eqn-momentum-conservation}). Equation~\ref{eqn-momentum-conservation} can then be written as,
\begin{equation}
\label{eq:hse2}
\left(\vturb^2 + \sum_{i=1,2}{\gamma_i K_i\rho^{\gamma_i-1}} \right) \rho^{-1}d\rho = - \frac{GM(r)dr}{r^2} ~~~.
\end{equation}
$K_i$ are constant with radius and are calculated at the boundary as functions of the gas properties - the temperature, $T_{{\rm th},b}$, density, $\rho_{b}$ and amount of non-thermal support. The latter is parameterized in \cite{Faerman2017} as $\alpha \equiv (P_{\rm th}+P_{\rm nth})/P_{\rm th}  = (T_{\rm th}+T_{\rm nth})/T_{\rm th}$~\footnote{~$\alpha$ is constant with radius for an isothermal gas distribution. 
In FSM20, the relative fractions of pressure support of each component vary with radius, and $\alpha$ is not constant.}. The entropy parameters are then given by
\begin{equation}\label{eq:k1}
K_1 = \frac{\kb}{\mbar^{\gamma_1} } \frac{T_{{\rm th},b}}{n_{b}^{\gamma_1-1} } ~~,~~
K_2 = \frac{\kb}{\mbar^{\gamma_2}} \frac{(\alpha_{b} - 1)T_{{\rm th},b}}{n_{b}^{\gamma_2-1}} ~~~,
\end{equation}
where $\alpha_{b} \equiv \alpha(\rcgm)$. Figure \ref{fig-illustration} (top-right) illustrates the rapidly decreasing gas temperature and increasing nonthermal contribution to total pressure (from turbulence, magnetic fields, and cosmic rays) with increasing galactocentric radius in the isentropic model.

To summarise, the model's input parameters are the gas density, temperature, ratio of nonthermal to thermal pressure at the halo boundary, and the turbulent velocity in the CGM. Setting these allows us to solve Equation~\ref{eq:hse2} for the gas density profile, $\rho(r)$, and then use the constant-entropy EoS (a solution to Equation~\ref{eqn-energy-conservation-2}) to find the pressure profiles for each of the components.

In FSM20, the CGM metallicity varies with the distance from the galaxy, and the gas ionisation state is set by collisional ionisation and photoionisation by the (redshift-dependent) metagalactic radiation field \citep{Haardt2012, Ferland2017}. In this work, to compare with other CGM models, we set the metallicity in the isentropic model to be constant with radius and assume only collisional ionisation equilibrium.

Given the distribution of gas and metals and the gas cooling function, the model calculates the radiative cooling luminosity of the CGM. These radiative losses can be translated to the mass cooling rate as a function of radius and integrated to give the global cooling rates for the entire corona. The model assumes that, on average, the CGM is in equilibrium, and these losses are balanced by energy inputs from processes such as galactic feedback, accretion, dissipation of turbulence, etc., or mass inputs from accretion and galactic outflows. This balance does not have to be perfect for star-forming galaxies and on short timescales, allowing for cooling-heating or accretion-outflow cycles (see the discussion in \citealt{Faerman2022}). In summary, the model requires a time-averaged net heating-cooling balance with star formation (i.e., $\left<\mathcal{C-H}\right> \propto \left<{\rm SFR}\right>$). \citet{FW2023} extend this model by adding a cool gas phase, in heating/cooling and ionisation equilibrium with the metagalactic radiation field, formed by precipitation from the hot phase, accreted from the IGM, stripped from satellites, or ejected from the galaxy.

\begin{table*}
\caption{A summary of the four CGM models. Here $t_{\rm cool}$ is gas cooling timescale, $t_{\rm ff}$ is free-fall timescale and $\rm M_{CGM}$ is the CGM mass.}
\begin{tabularx}{\textwidth}{c|*{4}{>{\raggedright\arraybackslash}X}}
 &  Precipitation & Isentropic & Cooling Flow & Baryon pasting \\
\midrule
\\
Momentum conservation & Hydrostatic equilibrium & Hydrostatic equilibrium & Hydrostatic equilibrium up to $(t_{\rm cool}/t_{\rm ff})^{-2}$ & Hydrostatic equilibrium \\\\
Functional constraint & Radially independent $t_{\rm cool}/t_{\rm ff}$ & Constant entropy & Energy + mass conservation in hot inflow & Polytropic relation \\\\
Boundary condition & Gas temperature at outer boundary &  Gas temperature at outer boundary & Gas temperature at outer boundary and SFR (or $\rm M_{CGM}$) & Confining pressure\\\\
Dynamical support & None & Constant turbulent velocity dispersion & Small, of order $(t_{\rm cool}/t_{\rm ff})^{-2}$ & Non-thermal pressure profile as parametric model with free parameters\\\\
Other support & None & Magnetic fields + cosmic rays as a relativistic fluid & None & None\\
\bottomrule
\end{tabularx}
\label{tab-input}
\end{table*}

\subsection{Cooling flow model}
\newcommand{\tcool}[0]{t_{\rm cool}}
\newcommand{\tff}[0]{t_{\rm ff}}
\newcommand{\tflow}[0]{t_{\rm flow}}
\newcommand{\Rcirc}[0]{R_{\rm circ}}
The cooling flow model discussed in \cite{Stern2019} assumes that the dynamical and heating effects of feedback on the CGM are small during the last cooling timescale $\tcool$. This assumption is expected to be valid either if feedback occurs in bursts that are separated by more than $\tcool$ or in low-redshift galaxies in which the effect of feedback heating on the CGM was strong at high redshift and has since subsided. This assumption is also more easily satisfied at small CGM radii, where $\tcool$ is a few $100\,{\rm Myr}-1\,{\rm Gyr}$, in contrast to large CGM radii where $\tcool$ can reach a Hubble time. It is thus plausible that the CGM forms a cooling flow at small radii where $\tcool$ is sufficiently short, while at large radii, the CGM more resembles the `thermal balance' models considered above.

At radii where the cooling flow assumption is satisfied, Equation~\ref{eqn-energy-conservation-2} implies,
\begin{equation}
    \frac{d\ln K}{d r} =  -\left[v_r\Bigl(\frac{P_{\rm th}}{\gamma_{\rm th}-1}\Bigr)\right]^{-1} \mathcal{C} = -\frac{1}{v_r\tcool}
\end{equation}
where for simplicity we assumed $P_{B} = P_{CR} = P_{turb}= 0$, and we used the definition of the cooling time $\tcool\equiv(\gamma-1)^{-1}P_{\rm th}/\mathcal{C}$. The model is illustrated in the bottom-left of Figure \ref{fig-illustration}. 
Note that despite the name `cooling flows', the inflowing gas remains hot down to the galaxy scale, since radiative losses are compensated by compressive heating. 

In the limits $\tcool^2\gg\tff^2$ (as expected in $\rm M_{vir}\gtrsim10^{12}\,{\rm M}_\odot$ halos) and $\mathcal{C}\gg\mathcal{H}$, 
analytic calculations and hydrodynamic simulations demonstrate that the CGM converges on a specific solution to Equations~\ref{eqn-mass-conservation}--\ref{eqn-energy-conservation-2} in which
\begin{equation}
    \frac{d\ln K}{d \ln r} = \frac{r/v_r}{\tcool}  \approx 1+\frac{4}{3}m
\end{equation}
where, $m\equiv d\ln v_{\rm c}/d\ln r$, and the approximation is exact for a power-law potential of the form $v_{\rm c}\propto r^m$. For an isothermal potential $m=0$, therefore, we get $K\propto r$. 
Also, since $v_r\approx r/\tcool$, we see that the ratio of the inertial term to the gravitational term in Equation~\ref{eqn-momentum-conservation} is of the order $(v_r/v_c)^2\approx (\tcool/\tff)^{-2}\ll1$, and hence a cooling flow is similar to the hydrostatic models considered above, with small deviations from hydrostatic equilibrium of the order $(\tcool/\tff)^{-2}$.

At small radii, centrifugal forces induced by angular momentum will change the structure of the hot gas and break its spherical symmetry. The implied axisymmetric solution was recently derived for the cooling flow model by \cite{Stern2023}, who showed that rotational support induces deviations from the hydrostatic equilibrium of order $(r/\Rcirc)^{-2}$, where the circularization radius $\Rcirc$ is defined through $j = v_c (\Rcirc )\Rcirc$ and depends on the spin of the inflowing hot gas. For MW-like halos, we expect $\Rcirc\approx15$ kpc. At $r \leq \Rcirc$, angular momentum support causes the hot inflow to halt, flatten into a disk geometry, cool and from $\sim10^6$ K to $\sim10^4$ K at the disc-halo interface. We discuss the implications of angular momentum on our results in Section \ref{sec-thermodynamic-profiles}.

\subsection{Baryon Pasting model}
\label{sec-bp-model}
The Baryon Pasting (BP) model is an analytic model of halo gas initially developed for galaxy clusters. It models gas with a polytropic equation of state, where the pressure of the gas is related to its density, including essential physics such as cooling, star formation, and feedback \citep{ostriker05}, non-thermal pressure due to bulk and turbulent gas motions \citep{Shaw2010}, cool-core \citep{{Flender2017}}, and gas density clumping \citep{shirasaki_etal20}. The latest BP model also features the painting of gas on DM particle in $N$-body simulation, in addition to painting gas on DM halo \citep{osato22}. 

The BP model assumes that the total pressure $P_{\rm{tot}}$ (thermal + nonthermal) of the halo gas is in hydrostatic equilibrium with the gravitational potential of the DM halo. The total pressure is related to the gas density through the polytropic relation: 
\begin{equation}
P_{\rm{tot}}(r) = P_0 \theta(r)^{n+1}
\end{equation}
where the gas density is given by $\rho_{g}(r) = \rho_0 \theta(r)^{n}$,  $\theta(r) = 1 + \frac{\Gamma-1}{\Gamma}\frac{\rho_0}{P_0}(\Phi_0 - \Phi(r))$ is a dimensionless function that represents the gas temperature, $\Phi_0$ is the central potential of the halo, and ${\Gamma=1+1/n}$ is the polytropic exponent, a parameter in the BP model. We set $\Gamma = 1.2$ outside cluster cores ($r> 0.2 R_{500c}$) as suggested from both cosmological hydrodynamical simulations and observations \citep[see e.g.,][]{Voit2005}. Within the core ($r < 0.2 R_{500c}$, by $\rm R_{break}$ in the bottom-right panel of Figure~\ref{fig-illustration}), the polytropic equation of state is modelled as $\Gamma_{\rm mod} = \Gamma_{\rm mod,0} (1+z)^{\beta}$, including the dependence on redshifts. 

The normalisation constants of the pressure and gas density profile, $P_0$ and $\rho_0$, respectively, are determined numerically by solving the energy and momentum conservation equations. In particular, the energy of the gas is given by
\begin{equation} \label{eq:E}
E_{g,f}  =  E_{g,i} + \epsilon_{DM} |E_{\rm DM }| + \epsilon_f M_\star c^2 + \Delta E_p.
\end{equation}
where $E_{g,f}$ and $E_{g,i}$ are the final and initial total (the sum of kinetic, thermal, and potential) energies of the gas. $\Delta E_p$ is the work done by the gas as it expands. $\epsilon_{DM} |E_{\rm DM}|$ is the energy transferred to the gas during major halo mergers through dynamical friction.\footnote{The exact value of $\epsilon_{DM}$ remains uncertain and is likely to depend on other factors, such as the merger history of a given halo.} 
The term $\epsilon_f M_\star c^2$ is the energy injected into the gas due to feedback from both supernovae (SNe) and AGN, and $M_\star$ is the total stellar mass. The slope and normalisation of the stellar mass-halo mass relation are two of the model's free parameters. Note that these two constraint equations are re-expressions of the conservation equations (Eqs.~\ref{eqn-mass-conservation},~\ref{eqn-momentum-conservation},~\ref{eqn-energy-conservation-2}).

The baryon pasting model includes the effects of non-thermal pressure in the gas by adopting the ``universal'' non-thermal pressure fraction profile \citep{Nelson2014}. The nonthermal pressure fraction is defined as $P_{\rm nt} = P_{\rm tot}(1-f_{\rm th})$.
The free parameters of the model are calibrated with the density profiles of the {\em Chandra}-SPT sample, covering a mass-limited cluster sample with $M_{500c} \geq 3\times 10^{14} M_\odot$ for redshift up to $z=1.7$, as well as the gas mass and total mass relations in clusters and groups from the {\em Chandra} and XMM-{\em Newton} data (see Section 3 and Table 3 of \citealt{Flender2017} for the details of the model parameters and their fiducial values). 

Note that the baryon basting model used here is an updated version of \citealt{Flender2017}. Instead of setting the pressure boundary to be the virial radius, the updated model uses $\rm R_{200m}$. This radius is a better match to the splash-back radius, which is closer to a physical boundary of the halo gas than the virial radius \citep[e.g.,][]{Shi2016,Aung2021}.


\vspace{0.5cm}


 We have built a Python pipeline (\url{https://github.com/psingh220/scam\_cgm}) to systematically compare different analytical CGM models. This pipeline allows for a fair comparison by properly assessing the quantities that are used in the predictions of the models. The CGM modelling interface provides a common platform for inputting parameters such as halo potential, metallicity profile, cooling function, boundary conditions, and model-specific parameters into individual CGM models. We will now present the output of our pipeline.

\begin{figure*}
\centering
	\includegraphics[height=13cm,angle=0.0 ]{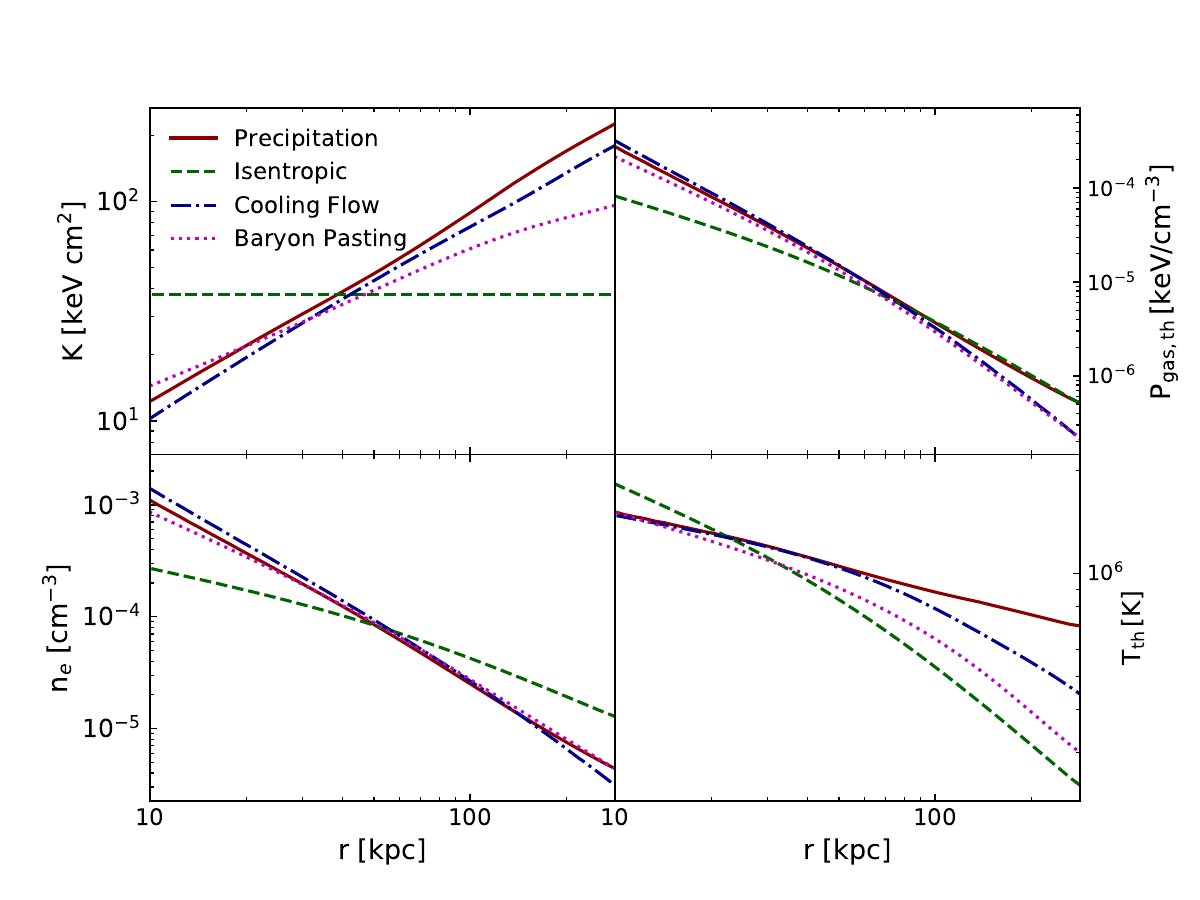}
	\caption{Specific entropy (top-left panel), thermal gas pressure (top-right panel), electron density (bottom left-hand panel) and temperature (bottom-right panel) profiles for a halo with $M_{\rm vir} \sim 10^{12} ~M_{\odot}$ at $z \sim 0$, for the precipitation (solid red), isentropic (dashed green), cooling flow (dashed-dotted blue) and baryon pasting (dotted magenta) models. Additional details on input quantities and fiducial model parameters are presented in Table~\ref{tab-common-parameters} and Section~\ref{sec-thermodynamic-profiles}.}
	\label{fig-tdQ-profiles}
\end{figure*}

\begin{table}
\caption{Common input parameters to the CGM models.}
\centering
\begin{tabular}{c | c}
\toprule
    $\rm M_{vir} $ & $  10^{12}~M_{\odot}$\\[1mm]
   $z$ & $0.001$\\
   Gravitational potential & NFW+galaxy with \\
   & concentration $c_{\rm vir}=10$\\[1mm]
   $M_* $ & $  6\times 10^{10}~M_{\odot}$\\[1mm]
   Disk scale length & 2.5 kpc\\[1mm]
   Metallicity & $ 0.3 {\rm Z}_{\odot}$ (uniform)\\[1mm]
   Ionisation \& Cooling & Collisional ionisation equilibrium\\[1mm]
  Outer boundary & $280\, \rm kpc$\\
  \bottomrule
\end{tabular}

\label{tab-common-parameters}
\end{table}

\section{Thermodynamic profiles of the CGM}
\label{sec-thermodynamic-profiles}
 
Table \ref{tab-input} summarises the modelling assumptions for the four models presented in Section \ref{sec-analytical-cgm}. The conservation of momentum is reduced to hydrostatic equilibrium in all four CGM models (in a cooling flow, hydrostatic equilibrium holds up to deviations of the order $\tcool^2/\tff^2$). The functional constraint signifies where the models' assumptions about CGM properties differ while solving Euler's equations. The isentropic and baryon pasting models include additional nonthermal pressure support in Equations~\ref{eqn-momentum-conservation} and \ref{eqn-energy-conservation-2} through turbulence, magnetic fields, and cosmic rays.

In Figure \ref{fig-tdQ-profiles}, we compare the gas entropy, thermal pressure, electron density, and temperature profiles for the four CGM models. The presented thermodynamic profiles are derived for a halo mass $\rm M_{vir} = 10^{12}~M_{\odot}$ at redshift $z\sim 0$, in an NFW+galaxy potential with concentration $\rm c_{vir}=10$, $\rm M_* = 6\times 10^{10}~M_{\odot}$ and a galactic disk scale length of 2.5 kpc. The cooling function is calculated using the {\tt Cloudy 17.00} \citep{Ferland2017} tables under collisional ionisation equilibrium (CIE) for uniform metallicity. For simplicity, we fix the CGM metallicity to $0.3~Z_{\odot}$. The boundary conditions are provided at 280~kpc (close to the virial radius) except for the baryon pasting model where it is 365~kpc (i.e. $\rm R_{200m}$).

Thermodynamic profiles for the precipitation model are shown for $t_{\rm cool}/t_{\rm ff}=10$ and gas temperature $\sim$ 0.06 keV ($\approx 7 \times 10^5$~K) at the boundary. The isentropic model profiles are shown for the fiducial model from FSM20, with $ T_{\rm b} = 2.4 \times 10^5$~K and $n_{\rm b} = 10^{-5}~{\rm cm^{-3}}$ (see Table 1). For the cooling flow model, we show a 1D non-rotating solution with mass inflow rate $\Dot{M}=1~{\rm M}_{\odot}\, {\rm yr}^{-1}$.  Table \ref{tab-common-parameters} summarises the common input parameters of the models. 

We note that angular momentum is expected to cause significant deviations from spherical symmetry at radii $\lesssim\Rcirc\sim15$ kpc. Specifically in the cooling flow solution, \cite{Stern2023} showed that this results in higher densities and lower temperatures in the disk plane versus the rotation axis and the spherically symmetric solution shown in Figure \ref{fig-tdQ-profiles}. The deviations scale as $(r/\Rcirc)^{-2}$, i.e.~they become rapidly stronger with decreasing radius. 

The slopes of the thermodynamic profiles in the isentropic model are considerably different from those of the other models. The gas temperature in the isentropic model is notably lower at $r \gtrsim 30$ kpc due to the lower boundary temperature adopted and the presence of non-thermal pressure support. The latter also results in a shallower density profile, leading to lower gas densities in the central region and higher densities at large radii. The fiducial precipitation and cooling flow models show similar thermodynamic profiles at $r \lesssim 100$~kpc. At larger radii, they differ in gas temperature and pressure, though at these large radii, $\tcool$ approaches the Hubble time, so it is not clear that a cooling flow has time to develop.

For the baryon pasting model, the profiles shown are for the best-fitting parameters fitted to the cluster and group X-ray observations from Table 3 of \cite{Flender2017} except for the feedback efficiency. The feedback energy from SN and AGN per stellar mass, $\epsilon_f = E_{\rm feedback}/(M_\star c^2)$, which is set in this work to $10^{-6}$, a factor of four lower than the fiducial value, $\epsilon_f=4\times10^{-6}$, which was calibrated from galaxy cluster observations. The higher feedback efficiency and, hence, the higher feedback energy that better describes ICM pushes more CGM outside the potential well. It also heats the gas to a higher temperature. This leads to a density and pressure that are more than an order of magnitude lower and a higher temperature for the baryon pasting model than other CGM models. Lowering the feedback efficiency to $10^{-6}$ brings the BP model into a much better agreement with the other CGM models. If the model is accurate, this suggests that the CGM favours a lower feedback efficiency than the ICM. This is consistent with the scenario in which SN feedback alone can provide enough energy to lift the CGM gas from the bottom of the potential well. However, agreement with other models alone does not mean that the real CGM prefers a lower feedback efficiency. For example, recent ACT SZ measurements of gas profiles of stacked massive galaxies and groups \citep{Amodeo2021} prefer higher feedback efficiencies. These observations suggest much higher thermal pressure and density profiles than state-of-the-art simulations.

The slope of the entropy profiles is $\sim 0.8-0.9$ for the precipitation and cooling flow models, $\sim 0.6$ for the baryon pasting model, and 0 (by construction) for the isentropic model. As noted earlier, the cooling flow and precipitation can be differentiated using a higher gas temperature at large galactocentric distances. Therefore, the combination of entropy and temperature profiles is the best discriminator among these models. Observational probes (or their combinations) that directly probe the entropy profile can be used to test the isentropic model. Precipitation-regulated gaseous halos have a distinctively higher entropy and temperature at the galaxy outskirts. Cooling flows lead to the steepest entropy, density, and pressure profiles. Consequently, we need radially resolved profiles and probes sensitive to large galactocentric distances to discern the dominant physical processes shaping the global properties of the CGM, and we now present predictions for some key observational probes.

\begin{figure*}
\centering
	\includegraphics[height=17cm,angle=0.0 ]{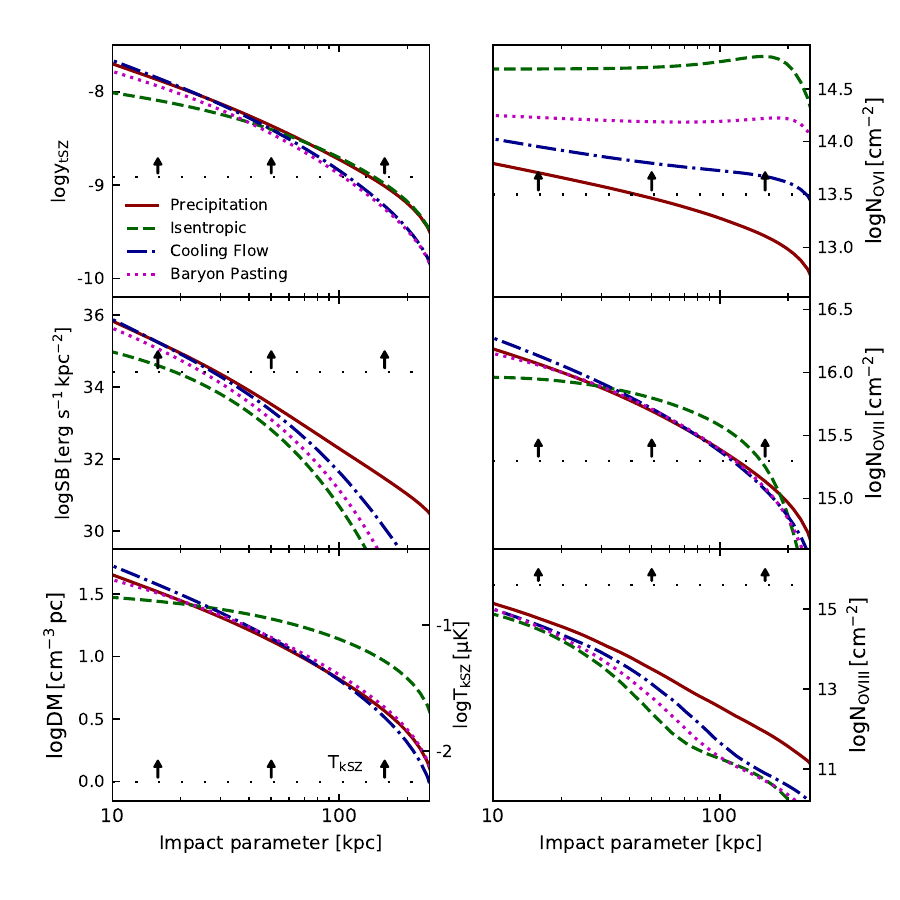}
	\caption{Comparison of the fiducial model's predictions for the projected tSZ effect (top left-hand panel), soft (0.5--2 keV) X-ray brightness (middle left-hand panel), dispersion measure and kSZ effect (bottom left-hand panel), and OVI, OVII, OVIII column densities (right-hand panels). Line styles and colours are as in Figure \ref{fig-tdQ-profiles}. The combination of X-ray emission and dispersion measures and OVII and OVIII absorption lines shows complementary trends and, therefore, can also be used to constrain the thermodynamics of the CGM. The loosely dotted lines (with arrows) correspond to the tSZ sensitivity limit for CMB-S4 for a stack of $\sim ~300,000$ galaxies at 150 GHz (top left-hand panel), eRASS4 sensitivity limit (radially averaged) for a stack of $\approx 7\times10^4$ isolated MW mass galaxies with median redshift $\approx 0.14$ from DESI Legacy survey (middle left-hand panel), kSZ sensitivity limit for CMB-S4 for a stack of $\sim ~100,000$ galaxies at 150 GHz (bottom left-hand panel), COS-Halos sensitivity limit for OVI (top-right panel), Athena/Arcus like mission sensitivity limits for OVII (middle-right panel) and OVIII (bottom-left panel) column densities for an individual absorber sightline.}
	\label{fig-Obs-profiles}
\end{figure*}

\section{CGM observables}
\label{sec-multi-wavelength}
In this section, we calculate the SZ effect, soft X-ray emission, FRB dispersion measure, and column densities of OVI, OVII, and OVIII, measurable in absorption, for the fiducial models described in Section \ref{sec-thermodynamic-profiles}. All these quantities are calculated and plotted for an external observer looking through the CGM. 

\subsection{Sunyaev-Zeldovich effect}
\label{sec-sz}
SZ effect is a secondary distortion in the black-body spectrum of the CMB through inverse Compton scattering of low-energy CMB photons with high-energy electrons present in the intervening ionised gaseous medium \citep{SZ1972}. 
The tSZ effect traces the thermal gas pressure and is characterised by a dimensionless $y$-parameter defined as
\begin{equation}
    y_{\rm tSZ} = \frac{\sigma_T}{m_e c^2}\int P_e \rm \,dl,
\end{equation}
where $\sigma_T$ is the Thompson scattering cross-section, $P_e$ is the electron pressure, and integration is performed along the line-of-sight. The change in CMB temperature due to the tSZ effect is a multiplication of the $y$-parameter and its unique frequency dependence. The kinetic SZ (kSZ) effect could be used to constrain the gas density modulo the line-of-sight bulk gas velocity and is directly given by the decrement caused by it in the CMB temperature.
\begin{equation}
    T_{\rm kSZ} = -T_{\rm CMB}\frac{\sigma_T}{c}\int n_e v_{los}\rm \, dl,
\end{equation}
where $n_e$ is the electron density, $v_{los}$ is the line-of-sight velocity of the medium away from us and $T_{\rm CMB}$ is the temperature of the un-distorted CMB.

In the top left-hand panel of Figure \ref{fig-Obs-profiles}, we show the projected $y$-parameter profiles due to the tSZ effect (or simply the tSZ profiles) for the four fiducial CGM models. The precipitation, cooling flow, and baryon pasting models predict very similar tSZ profiles at small scales ($r<50$ kpc, although note the angular momentum effects for the cooling flow model discussed above). In contrast, the precipitation and isentropic models converge at galaxy outskirts, tracing their pressure profiles. The cooling-flow and baryon pasting models predict a comparatively steeper, and the isentropic model predicts a shallower tSZ profile. The fiducial model predictions differ by a maximum factor of two near the galaxy centre and outskirt, while converging at intermediate scales. These results indicate the need for higher-angular-resolution CMB experiments probing smaller scales to use the tSZ signal to differentiate the CGM models.

Stage-3 (AdvACT and SPT-3G) and Stage-4 (CMB-S4) CMB surveys will play a critical role in resolving SZ profiles with high signal-to-noise ratio, providing the opportunity to constrain CGM physics down to $10^{12}~\rm M_{\odot}$ \citep{Battaglia2019}. The dotted line and arrows show the expected sensitivity limit $\sim 10^{-9}$ for CMB-S4 at 150 GHz (assuming the noise RMS of $1.8 \rm \mu K$-arcmin, see Table~1 of \citealt{Battaglia2017}) for a stack of $\approx 300,000$ galaxies. The same frequency channel is expected to have an angular resolution of $\sim$ 1 arcmin, corresponding to a spatial resolution of 50--100 kpc for a median redshift of $\sim 0.05-0.1$. Combining several frequency channels would significantly improve the sensitivity limit, at the cost of lower angular resolution.

We show the kSZ profile predictions in the bottom-left panel (right y-axis), assuming a typical peculiar velocity of $\rm 300 \,km\, s^{-1}$ \citep{Schaan2021, Tanimura22}. The shape of the kSZ profile is identical to the FRB dispersion measure profile (Section \ref{sec-frb} below) since both observables trace the electron density integrated along the line of sight\footnote{except if significant rotation is present, see \cite{Stern2023}.}. Note that the change in CMB temperature induced by the kSZ signal is larger than that of the tSZ signal ($=y_{\rm tSZ}T_{\rm CMB}$) throughout an $\rm M_{vir} \sim 10^{12}~M_{\odot}$ halo, in contrast to galaxy clusters ($\rm M_{vir} \sim 10^{15}~M_{\odot}$) where the tSZ signal dominates due to the higher ICM temperature. The ratio of kSZ to tSZ decrements is lowest in the precipitation model (kSZ/tSZ $\sim$ 3 and increases with increasing distance from the centre) in the low-frequency limit (where the tSZ frequency-dependent factor $\approx -2$). The ratio increases to $\sim 5$ for the cooling flow and to $\sim 10$ for the BP and isentropic models. 

The kSZ sensitivity limit of CMB-S4 at 150 GHz is an order of magnitude above the predictions for the four fiducial models. Assuming that an accurate galaxy peculiar velocity estimation can be obtained with an overlapping spectroscopic survey like DESI \citep{Ried2023}, a sensitivity of $\sim 5\times10^{-3}\, \mu {\rm K}$ can be achieved by stacking $\sim 100,000$ galaxies. Such sensitivity levels are sufficient to detect the kSZ signal out to the virial radius for our fiducial models. Therefore, extracting accurate galaxy peculiar velocities poses both a challenge and an exciting avenue for studying the CGM in $L_*$ galaxies with the kSZ effect in future CMB surveys and advanced techniques.

\subsection{Soft X-ray emission}
Detection of extended X-ray emission from nearby galaxies is one of the few direct observations of hot CGM \citep{Anderson2011, Anderson2016, Bogdan2017, Li2018}. The emission is highly sensitive to the gas density ($\propto n^2_e$) and provides a valuable probe of the CGM distribution in galaxies. 

The middle-left panel of Figure \ref{fig-Obs-profiles} compares the soft X-ray ($0.5-2$~keV) surface brightness profiles predicted by the models. Note that the X-ray emission in a given energy band falls rapidly as the gas temperature moves away from the band. If the gas temperature is within the energy band considered, the X-ray emission is primarily dictated by the gas density. As a result, the model predictions for the X-ray surface brightness profiles trace the gas temperature at large radii and transition to tracing the gas density at smaller radii. The isentropic model gives the faintest X-ray halo due to low gas densities at small radii and low gas temperatures at large radii (see Figure \ref{fig-tdQ-profiles}). The baryon pasting and cooling-flow models show similar X-ray emission profiles throughout the halo. The precipitation and cooling flow models predict similar X-ray emission out to 50~kpc, beyond which the precipitation model takes over, thus predicting the brightest X-ray halos within $R_{\rm vir}$. Therefore, measurements of X-ray surface brightness profiles near the virial radius (where the CGM densities are low) can be used to put stringent constraints on gas temperature and the thermal versus non-thermal components of a given CGM model. 

However, measurement of the surface brightness beyond a few tens of kpc has been challenging because of the rapid signal decline with decreasing density. In the same panel, we show the average (over radial bins) sensitivity level for eRASS4+DESI Legacy survey \citep{Zhang2024} by stacking a sample of $\approx 7\times10^4$ isolated MW mass (median $M_* \approx 5.5\times 10^{10} M_\odot$) galaxies with the median $z\leq 0.14$. For this sensitivity limit and our fiducial model parameters, the X-ray emission is detectable barely out to 20--30\,kpc from the halo centre. The predicted X-ray emission profiles are sensitive to assumed virial mass and model-specific parameters. For example, \cite{Chadayammuri2022} measured the surface brightness at $\gtrsim$ 100\,kpc by stacking 2643 galaxies in the X-ray emission from the eFEDS survey \citep{Brunner2022}, and the stacked signal is dominated by brighter (and hence more massive than our fiducial model) galaxies. Furthermore, at temperatures of $\sim 10^6$\,K, expected for the CGM of $L_*$ galaxies, the emission in the soft X-ray is dominated by metal lines, and the degeneracy between gas metallicity and temperature limits the power of X-ray emission alone to simultaneously constrain the two \citep{Anderson2013, Das2021}.

\subsection{Fast Radio Bursts}
\label{sec-frb}

The impact of the intervening ionised medium on the FRB signal causes a frequency-dependent delay in its arrival, represented by the dispersion measure (DM) \citep[e.g.,][]{McQuinn2014,Ravi2019,Chawla2022}. The DM is thus sensitive to the ionised CGM and its dependence on feedback physics \citep[e.g.,][]{Medlock2024}. DM predictions for our models are plotted in the bottom-left panel of Figure~\ref{fig-Obs-profiles}. The isentropic model also has a high DM due to the high electron density at large radii, resulting from nonthermal pressure. The cooling flow, precipitation, and baryon pasting models are consistent with each other throughout the radial range. At large radii, near $r \sim 100-150$~kpc (roughly half of the virial radius of the galaxy), the DM of the isentropic model is a factor of $\sim 2$ higher than compared to the predictions for the other models. The isentropic model also predicts a flatter DM profile compared to the others, due to its flatter density profile. As noted above, the kSZ profile is identical to the DM signal (neglecting CGM rotation effects).

Presently available measurements of the DM are limited to more massive galaxies \citep{Connor2022, Wu2022} or only upper limits ($\rm \sim 100 \,cm^{-3}\,pc$ from \citealt{Ravi2023} and $\rm \sim 200\, cm^{-3}\,pc$ from \citealt{Cook2023} for the MW CGM for an external observer at the solar circle), and are consistent with all four fiducial CGM models. Therefore, they cannot pinpoint CGM thermodynamics in $L_*$ galaxies. We note that the measurement of DM from MW mass galaxies is not limited by the sensitivity limits of the corresponding missions (which are not shown in the figure) but by the limited spatial localisation of FRB sightlines. The uncertainties in the contribution of the CGM to the total measured DM can only be improved with better FRB localisations and a large statistical sample of FRBs \citep{Scott2023, Jankowski2023}.

\subsection{Absorption lines}\label{sec-absorption}

Absorption line studies provide some of the most stringent constraints on the CGM mass in different phases, metallicity, ionisation state, and the extent of the gas \citep{Werk2014, Johnson2015, McQuinn2018}. High ions such as OVI, OVII, OVIII, NeVIII, and FeXVII observed at UV and X-ray wavelengths are particularly useful for constraining warm/hot CGM \citep{Bregman2007, Tumlinson2011, Gupta2012, Burchett2019, Tchernyshyov2022, Qu2024}.

In the right panels of Figure \ref{fig-Obs-profiles}, we show the predictions of the fiducial model for the column densities of OVI (top), OVII (middle) and OVIII (bottom). These column densities depend on the product of gas density, metallicity, and ion fraction, where the ion fractions themselves depend on the gas properties and assumptions about the ionisation mechanism. Throughout this work, we calculate the ion fractions using {\tt Cloudy 17.00} \citep{Ferland2017} assuming an optically thin gas in CIE, with the ion fractions functions of the gas temperature alone\footnote{We note that at low gas densities in the CGM, photoionisation by the metagalactic radiation field can be significant even for the high ions discussed here and at low redshift (see discussion in FSM20 and \citealp{Faerman2022}).}.

The OVI ion fraction peaks at temperatures $\sim 3\times 10^5$~K in CIE. In the isentropic model, the gas temperature at large radii is close to this value, resulting in a high $N_{\rm OVI}$. As the CGM temperature at $r>100$ kpc in the precipitation model is a factor of $\sim 2-3$ higher than in the isentropic model, $N_{\rm OVI}$ decreases by more than an order of magnitude throughout the halo (see also \citealt{voit2019_column_densities} for the implications of temperature fluctuations on the density of the OVI column). $N_{\rm OVI}$ predictions for the cooling flow and baryon pasting models lie in the range enclosed by the isentropic and precipitation models due to their intermediate gas temperatures. 

The OVI column density detection limit for the COS-Halos survey is $\geq 3 \times 10^{13}\, \rm cm^{-2}$ \citep{Tumlinson2011}, well below most of our predictions of the fiducial model, except at $r>40$ kpc (see also Appendix \ref{apn-model-unc} for a compilation of $N_{\rm OVI}$ measurements). However, the models discussed here address the warm/hot phase of the CGM and do not include the cool- or intermediate-temperature gas. Previous work explored different scenarios to understand whether the OVI can originate in this lower-temperature gas. For example, \citet{Stern2018} suggested that it may reside in low-density cool photo-ionised gas outside the virial shock. Another suggestion was that the observed OVI may form in gas cooling from the ambient hot phase or in mixing layers between the hot and cool phases. However, \citet{Gnat2004} showed that many interfaces are required to reproduce the OVI columns measured in the MW. Furthermore, \citet{FW2023} showed that even a significant mass of intermediate temperature gas (similar to the cool gas mass) would only contribute $\lesssim 10\%$ to the OVI column measured in the COS-Halos survey at large impact parameters.

The ionisation fraction of OVII remains relatively constant and close to unity at temperatures $T \sim 3\times 10^5-2\times 10^6$~K and falls rapidly outside of this range under the assumption of CIE. As a result, $N_{\rm OVII}$ approximately follows the respective gas density profiles, with the isentropic model predicting the highest column densities beyond 30 kpc and out to the virial radius. On the other hand, the ionisation fraction of OVIII peaks at $T \sim 2\times 10^6$~K. Therefore, the precipitation model predicts the largest $N_{\rm OVIII}$. The isentropic, cooling flow, and baryon pasting models predict large values of $N_{\rm OVIII}$ at $r<30$~kpc where the temperature is favourable for OVIII, followed by a rapid decline.

Currently, for OVII and OVIII, we are limited to column densities greater than $10^{16}$ and $2\times10^{15} \, \rm cm^{-2}$, respectively, detected in absorption by the MW CGM \citep{Bregman2007, Gupta2012, Fang2015, Das2019}. An Arcus \citep{Smith2016} or Athena \citep{Barret2016} like mission will be able to detect OVII column densities down to $\geq 2 \times 10^{15}\, \rm cm^{-2}$ for individual galaxies \citep{Wijers2020}. The sensitivity limit is shown in the middle-right panel of Figure \ref{fig-Obs-profiles}. Therefore, such missions will enable us to detect OVII in MW-like galaxies out to the virial radii and constrain the gas density distributions approximately independently of the temperature profile. LEM \citep{LEM2022} will allow us to probe these lines in X-ray emission beyond the virial radii for individual MW mass galaxies. Although the OVIII column density detection limits for these missions ($\geq 4 \times 10^{15}\, \rm cm^{-2}$) are higher than the fiducial model predictions, the sensitivity limit can be reduced by an order of magnitude or more by stacking a large number of galaxies.

In Appendix \ref{apn-model-unc}, we provide a qualitative comparison of our model predictions with currently available datasets for $L_*$ and massive galaxies, including model parameter uncertainties. Our publicly available Python toolkit\footnote{\url{https://github.com/ethlau/cgm_toolkit}} allows us to compute these observables, provided the CGM density, temperature, and metallicity profiles.

\section{Summary}
\label{sec-summary}
Idealised CGM models provide an intuitive method to constrain physical processes in the CGM. In this paper, we compare the thermodynamic profiles of the CGM for the precipitation, isentropic, cooling flow, and baryon pasting models. Each CGM model considered here solves Euler's fluid equations under model-specific assumptions (see Figure~\ref{fig-illustration} for illustration). We computed the entropy, electron density, gas pressure, and temperature profiles (Figure~\ref{fig-tdQ-profiles}) for the four fiducial models for a MW mass galaxy ($\rm M_{vir} = 10^{12}~M_{\odot}$, $z\sim 0$) in an NFW+galaxy potential with concentration $\rm c_{vir}=10$, $\rm M_* = 6\times 10^{10}~M_{\odot}$. We assume a gas in the CIE with a constant metallicity of $0.3~Z_{\odot}$. The models show the most significant differences in the entropy and temperature profiles at small ($r\lesssim 30$~kpc) and large ($r\gtrsim 100$~kpc) galactocentric distances. Specifically, the precipitation model predicts comparatively high entropy, high temperature ($T\sim 10^6$~K), and low CGM density beyond $\gtrsim 50$ kpc. The predictions of the cooling flow model are close to those of the precipitation model for the radii within the cooling radius ($r\lesssim 100$~kpc) where it is potentially valid. The isentropic model predicts relatively flatter profiles (except temperature), while the baryon pasting model shows a distinctively cored temperature profile at $r<30$ kpc.

We then addressed the observable quantities predicted by the fiducial models and compared the SZ effect, soft X-ray emission, DM probed by FRBs, and oxygen column densities for OVI, OVII and OVIII, measured in UV and X-ray absorption (Figure~\ref{fig-Obs-profiles}). These observables trace different combinations of the thermodynamic quantities, and their combinations can be used to constrain the underlying CGM physical properties. We used the same set of input parameters for each model, thus making sure that any differences in the predicted thermodynamic or observable quantities are due to the inherent differences in the model assumptions. 

The tSZ profiles trace the projected pressure profiles, differing by a factor of two near the galactic centre and its outskirt. Stacking $\sim 300,000$ galaxies with CMB-S4 could detect and resolve the tSZ signature predicted by the four fiducial models up to 100\,kpc. In the case of the kSZ effect, the sensitivity level required to probe the signal out to the virial radius can be achieved by stacking $\sim 100,000$ galaxies in CMB-S4 provided accurate peculiar velocity measurements. OVII column density traces the gas density profile in the temperature range $T \sim 3\times 10^5-2\times 10^6$\,K, which, in fact, captures the variety of temperature profiles predicted by these distinct models. The OVII column density signal predicted by the models can be detected out to 100\,kpc for individual galaxies for Athena/Arcus-like missions. The soft X-ray emission shows distinct predictions for the precipitation and the isentropic models, where the former produces at least an order of magnitude brighter X-ray halos than the latter. 
The current model predictions for the X-ray surface brightness are lower than the eROSITA detection limits, highlighting the need for deeper X-ray surveys.
The models also predict unique FRB DM profiles, where the combination of DM and soft X-ray emission displays the ability to differentiate between the precipitation and the isentropic models. Compared to the other models, the cooling flow model predicts steeper observable profiles in most cases (except for OVIII). Our results show that the amplitude of the kSZ effect is dominant (for typical peculiar velocities of $\rm \approx 300\, km\, s^{-1}$) compared to that of the tSZ effect in MW mass galaxies for all four CGM models (although it is more challenging to separate the kSZ signal from CMB fluctuations since there is no associated change in the spectrum). OVI column density predictions also show stark differences between the CGM models, although potential contributions from additional phases may complicate the interpretation of this observable (see Section~\ref{sec-absorption}).

We showed that combining entropy and temperature profiles allows one to assess the relative importance of the physical processes in the different CGM models. Still, we need a combination of observables to do so.  For example, tSZ alone is not a good diagnostic of CGM physics due to the similar predictions by very different models. However, when combined with the FRB dispersion measure or the kSZ signal, they can constrain both the temperature and entropy of the CGM ($K(r) \propto \rm tSZ/DM^{5/3}$ and $T(r) \propto \rm tSZ/DM$). Similarly, the combination of the OVII and OVIII column densities is particularly useful for simultaneously constraining the density and temperature of the CGM due to the sensitivity of the ion fractions to the gas temperature.

The Python package we used for computing the observables from the CGM models is made publicly available. It can be used easily to add another CGM model to this comparison with the code. It presents a unique platform for forward modelling the CGM observables from its thermodynamic properties.

There are several ongoing and next-generation missions planned with the CGM as one of the key science goals, with observations across the electromagnetic spectrum. In this work, we bring different idealised CGM models onto a single platform, and the accompanying pipeline will be useful for modelling and interpreting a plethora of multi-wavelength CGM observations, allowing a direct comparison of their data sets with a variety of easily tunable CGM physics realisations. Such an analysis is also essential to harness the capabilities of next-generation multi-wavelength missions as a community by differentiating CGM models with strikingly different assumptions on CGM physics. \\\\
{\bf{ACKNOWLEDGEMENTS}}\\
We thank the referee for valuable suggestions and comments. We thank Mark G. Voit, Benjamin D. Oppenheimer, Nicholas Battaglia, Yi Zhang, Srinivasan Raghunathan and Zhijie Qu for the helpful discussions. We gratefully acknowledge the hospitality and discussions at ``Fundamentals of Gaseous Halos" conference at the Kavli Institute for Theoretical Physics (KITP), which was supported in part by the National Science Foundation (NSF) under Grant No.~NSF PHY-1748958. PS acknowledges support from the YCAA Prize Postdoctoral Fellowship and the Yale Center for Research Computing facilities and staff. YF was supported by NASA award 19-ATP19-0023 and NSF award AST-2007012. JS was supported by the Israel Science Foundation (grant No. 2584/21). DN was supported by the NSF award AST 2206055.\\\\ 
{\bf{DATA AVAILABILITY}}\\
	The data sets underlying this article are publicly available or available via the authors of their respective publications.

	\footnotesize{
		\bibliography{references}{}

\begin{thebibliography}{96}
\expandafter\ifx\csname natexlab\endcsname\relax\def\natexlab#1{#1}\fi

\bibitem[{{Abazajian} {et~al}\mbox{.}(2016){Abazajian}, {Adshead}, {Ahmed}, {Allen}, {Alonso}, {Arnold}, {Baccigalupi}, {Bartlett}, {Battaglia}, {Benson}, {Bischoff}, {Borrill}, {Buza}, {Calabrese}, {Caldwell}, {Carlstrom}, {Chang}, {Crawford}, {Cyr-Racine}, {De Bernardis}, {de Haan}, {di Serego Alighieri}, {Dunkley}, {Dvorkin}, {Errard}, {Fabbian}, {Feeney}, {Ferraro}, {Filippini}, {Flauger}, {Fuller}, {Gluscevic}, {Green}, {Grin}, {Grohs}, {Henning}, {Hill}, {Hlozek}, {Holder}, {Holzapfel}, {Hu}, {Huffenberger}, {Keskitalo}, {Knox}, {Kosowsky}, {Kovac}, {Kovetz}, {Kuo}, {Kusaka}, {Le Jeune}, {Lee}, {Lilley}, {Loverde}, {Madhavacheril}, {Mantz}, {Marsh}, {McMahon}, {Meerburg}, {Meyers}, {Miller}, {Munoz}, {Nguyen}, {Niemack}, {Peloso}, {Peloton}, {Pogosian}, {Pryke}, {Raveri}, {Reichardt}, {Rocha}, {Rotti}, {Schaan}, {Schmittfull}, {Scott}, {Sehgal}, {Shandera}, {Sherwin}, {Smith}, {Sorbo}, {Starkman}, {Story}, {van Engelen}, {Vieira}, {Watson}, {Whitehorn}, \& {Kimmy Wu}}]{Abazajian2016}
{Abazajian} K.~N. {et~al.}, 2016, arXiv e-prints, arXiv:1610.02743

\bibitem[{{Ade} {et~al}\mbox{.}(2019){Ade}, {Aguirre}, {Ahmed}, {Aiola}, {Ali}, {Alonso}, {Alvarez}, {Arnold}, {Ashton}, {Austermann}, {Awan}, {Baccigalupi}, {Baildon}, {Barron}, {Battaglia}, {Battye}, {Baxter}, {Bazarko}, {Beall}, {Bean}, {Beck}, {Beckman}, {Beringue}, {Bianchini}, {Boada}, {Boettger}, {Bond}, {Borrill}, {Brown}, {Bruno}, {Bryan}, {Calabrese}, {Calafut}, {Calisse}, {Carron}, {Challinor}, {Chesmore}, {Chinone}, {Chluba}, {Cho}, {Choi}, {Coppi}, {Cothard}, {Coughlin}, {Crichton}, {Crowley}, {Crowley}, {Cukierman}, {D'Ewart}, {D{\"u}nner}, {de Haan}, {Devlin}, {Dicker}, {Didier}, {Dobbs}, {Dober}, {Duell}, {Duff}, {Duivenvoorden}, {Dunkley}, {Dusatko}, {Errard}, {Fabbian}, {Feeney}, {Ferraro}, {Flux{\`a}}, {Freese}, {Frisch}, {Frolov}, {Fuller}, {Fuzia}, {Galitzki}, {Gallardo}, {Tomas Galvez Ghersi}, {Gao}, {Gawiser}, {Gerbino}, {Gluscevic}, {Goeckner-Wald}, {Golec}, {Gordon}, {Gralla}, {Green}, {Grigorian}, {Groh}, {Groppi}, {Guan}, {Gudmundsson}, {Han}, {Hargrave}, {Hasegawa}, {Hasselfield},
  {Hattori}, {Haynes}, {Hazumi}, {He}, {Healy}, {Henderson}, {Hervias-Caimapo}, {Hill}, {Hill}, {Hilton}, {Hilton}, {Hincks}, {Hinshaw}, {Hlo{\v{z}}ek}, {Ho}, {Ho}, {Howe}, {Huang}, {Hubmayr}, {Huffenberger}, {Hughes}, {Ijjas}, {Ikape}, {Irwin}, {Jaffe}, {Jain}, {Jeong}, {Kaneko}, {Karpel}, {Katayama}, {Keating}, {Kernasovskiy}, {Keskitalo}, {Kisner}, {Kiuchi}, {Klein}, {Knowles}, {Koopman}, {Kosowsky}, {Krachmalnicoff}, {Kuenstner}, {Kuo}, {Kusaka}, {Lashner}, {Lee}, {Lee}, {Leon}, {Leung}, {Lewis}, {Li}, {Li}, {Limon}, {Linder}, {Lopez-Caraballo}, {Louis}, {Lowry}, {Lungu}, {Madhavacheril}, {Mak}, {Maldonado}, {Mani}, {Mates}, {Matsuda}, {Maurin}, {Mauskopf}, {May}, {McCallum}, {McKenney}, {McMahon}, {Meerburg}, {Meyers}, {Miller}, {Mirmelstein}, {Moodley}, {Munchmeyer}, {Munson}, {Naess}, {Nati}, {Navaroli}, {Newburgh}, {Nguyen}, {Niemack}, {Nishino}, {Orlowski-Scherer}, {Page}, {Partridge}, {Peloton}, {Perrotta}, {Piccirillo}, {Pisano}, {Poletti}, {Puddu}, {Puglisi}, {Raum}, {Reichardt}, {Remazeilles},
  {Rephaeli}, {Riechers}, {Rojas}, {Roy}, {Sadeh}, {Sakurai}, {Salatino}, {Sathyanarayana Rao}, {Schaan}, {Schmittfull}, {Sehgal}, {Seibert}, {Seljak}, {Sherwin}, {Shimon}, {Sierra}, {Sievers}, {Sikhosana}, {Silva-Feaver}, {Simon}, {Sinclair}, {Siritanasak}, {Smith}, {Smith}, {Spergel}, {Staggs}, {Stein}, {Stevens}, {Stompor}, {Suzuki}, {Tajima}, {Takakura}, {Teply}, {Thomas}, {Thorne}, {Thornton}, {Trac}, {Tsai}, {Tucker}, {Ullom}, {Vagnozzi}, {van Engelen}, {Van Lanen}, {Van Winkle}, {Vavagiakis}, {Verg{\`e}s}, {Vissers}, {Wagoner}, {Walker}, {Ward}, {Westbrook}, {Whitehorn}, {Williams}, {Williams}, {Wollack}, {Xu}, {Yu}, {Yu}, {Zago}, {Zhang}, {Zhu}, \& {Simons Observatory Collaboration}}]{Ade2019}
{Ade} P. {et~al.}, 2019, \jcap, 2019, 056

\bibitem[{{Amodeo} {et~al}\mbox{.}(2021){Amodeo}, {Battaglia}, {Schaan}, {Ferraro}, {Moser}, {Aiola}, {Austermann}, {Beall}, {Bean}, {Becker}, {Bond}, {Calabrese}, {Calafut}, {Choi}, {Denison}, {Devlin}, {Duff}, {Duivenvoorden}, {Dunkley}, {D{\"u}nner}, {Gallardo}, {Hall}, {Han}, {Hill}, {Hilton}, {Hilton}, {Hlo{\v{z}}ek}, {Hubmayr}, {Huffenberger}, {Hughes}, {Koopman}, {MacInnis}, {McMahon}, {Madhavacheril}, {Moodley}, {Mroczkowski}, {Naess}, {Nati}, {Newburgh}, {Niemack}, {Page}, {Partridge}, {Schillaci}, {Sehgal}, {Sif{\'o}n}, {Spergel}, {Staggs}, {Storer}, {Ullom}, {Vale}, {van Engelen}, {Van Lanen}, {Vavagiakis}, {Wollack}, \& {Xu}}]{Amodeo2021}
{Amodeo} S. {et~al.}, 2021, \prd, 103, 063514

\bibitem[{{Anderson} \& {Bregman}(2011)}]{Anderson2011}
{Anderson} M.~E., {Bregman} J.~N., 2011, \apj, 737, 22

\bibitem[{{Anderson}, {Bregman} \& {Dai}(2013){Anderson}, {Bregman}, \& {Dai}}]{Anderson2013}
{Anderson} M.~E., {Bregman} J.~N., {Dai} X., 2013, \apj, 762, 106

\bibitem[{{Anderson}, {Churazov} \& {Bregman}(2016){Anderson}, {Churazov}, \& {Bregman}}]{Anderson2016}
{Anderson} M.~E., {Churazov} E., {Bregman} J.~N., 2016, \mnras, 455, 227

\bibitem[{{Aung}, {Nagai} \& {Lau}(2021){Aung}, {Nagai}, \& {Lau}}]{Aung2021}
{Aung} H., {Nagai} D., {Lau} E.~T., 2021, \mnras, 508, 2071

\bibitem[{{Barret} {et~al}\mbox{.}(2016){Barret}, {Lam Trong}, {den Herder}, {Piro}, {Barcons}, {Huovelin}, {Kelley}, {Mas-Hesse}, {Mitsuda}, {Paltani}, {Rauw}, {Ro{\.Z}anska}, {Wilms}, {Barbera}, {Bozzo}, {Ceballos}, {Charles}, {Decourchelle}, {den Hartog}, {Duval}, {Fiore}, {Gatti}, {Goldwurm}, {Jackson}, {Jonker}, {Kilbourne}, {Macculi}, {Mendez}, {Molendi}, {Orleanski}, {Pajot}, {Pointecouteau}, {Porter}, {Pratt}, {Pr{\^e}le}, {Ravera}, {Renotte}, {Schaye}, {Shinozaki}, {Valenziano}, {Vink}, {Webb}, {Yamasaki}, {Delcelier-Douchin}, {Le Du}, {Mesnager}, {Pradines}, {Branduardi-Raymont}, {Dadina}, {Finoguenov}, {Fukazawa}, {Janiuk}, {Miller}, {Naz{\'e}}, {Nicastro}, {Sciortino}, {Torrejon}, {Geoffray}, {Hernandez}, {Luno}, {Peille}, {Andr{\'e}}, {Daniel}, {Etcheverry}, {Gloaguen}, {Hassin}, {Hervet}, {Maussang}, {Moueza}, {Paillet}, {Vella}, {Campos Garrido}, {Damery}, {Panem}, {Panh}, {Bandler}, {Biffi}, {Boyce}, {Cl{\'e}net}, {DiPirro}, {Jamotton}, {Lotti}, {Schwander}, {Smith}, {van Leeuwen}, {van
  Weers}, {Brand}, {Cobo}, {Dauser}, {de Plaa}, \& {Cucchetti}}]{Barret2016}
{Barret} D. {et~al.}, 2016, in Society of Photo-Optical Instrumentation Engineers (SPIE) Conference Series, Vol. 9905, Space Telescopes and Instrumentation 2016: Ultraviolet to Gamma Ray, {den Herder} J.-W.~A., {Takahashi} T., {Bautz} M., eds., p. 99052F

\bibitem[{{Battaglia} {et~al}\mbox{.}(2017){Battaglia}, {Ferraro}, {Schaan}, \& {Spergel}}]{Battaglia2017}
{Battaglia} N., {Ferraro} S., {Schaan} E., {Spergel} D.~N., 2017, \jcap, 2017, 040

\bibitem[{{Battaglia} {et~al}\mbox{.}(2019){Battaglia}, {Hill}, {Amodeo}, {Bartlett}, {Basu}, {Erler}, {Ferraro}, {Hernquist}, {Madhavacheril}, {McQuinn}, {Mroczkowski}, {Nagai}, {Schaan}, {Somerville}, {Sunyaev}, {Vogelsberger}, \& {Werk}}]{Battaglia2019}
{Battaglia} N. {et~al.}, 2019, \baas, 51, 297

\bibitem[{{Benson} {et~al}\mbox{.}(2014){Benson}, {Ade}, {Ahmed}, {Allen}, {Arnold}, {Austermann}, {Bender}, {Bleem}, {Carlstrom}, {Chang}, {Cho}, {Cliche}, {Crawford}, {Cukierman}, {de Haan}, {Dobbs}, {Dutcher}, {Everett}, {Gilbert}, {Halverson}, {Hanson}, {Harrington}, {Hattori}, {Henning}, {Hilton}, {Holder}, {Holzapfel}, {Irwin}, {Keisler}, {Knox}, {Kubik}, {Kuo}, {Lee}, {Leitch}, {Li}, {McDonald}, {Meyer}, {Montgomery}, {Myers}, {Natoli}, {Nguyen}, {Novosad}, {Padin}, {Pan}, {Pearson}, {Reichardt}, {Ruhl}, {Saliwanchik}, {Simard}, {Smecher}, {Sayre}, {Shirokoff}, {Stark}, {Story}, {Suzuki}, {Thompson}, {Tucker}, {Vanderlinde}, {Vieira}, {Vikhlinin}, {Wang}, {Yefremenko}, \& {Yoon}}]{Benson2014}
{Benson} B.~A. {et~al.}, 2014, in Society of Photo-Optical Instrumentation Engineers (SPIE) Conference Series, Vol. 9153, Millimeter, Submillimeter, and Far-Infrared Detectors and Instrumentation for Astronomy VII, {Holland} W.~S., {Zmuidzinas} J., eds., p. 91531P

\bibitem[{{Bogd{\'a}n} {et~al}\mbox{.}(2017){Bogd{\'a}n}, {Bourdin}, {Forman}, {Kraft}, {Vogelsberger}, {Hernquist}, \& {Springel}}]{Bogdan2017}
{Bogd{\'a}n} {\'A}., {Bourdin} H., {Forman} W.~R., {Kraft} R.~P., {Vogelsberger} M., {Hernquist} L., {Springel} V., 2017, \apj, 850, 98

\bibitem[{{Bregman} {et~al}\mbox{.}(2022){Bregman}, {Hodges-Kluck}, {Qu}, {Pratt}, {Li}, \& {Yun}}]{Bregman2022}
{Bregman} J.~N., {Hodges-Kluck} E., {Qu} Z., {Pratt} C., {Li} J.-T., {Yun} Y., 2022, \apj, 928, 14

\bibitem[{{Bregman} \& {Lloyd-Davies}(2007)}]{Bregman2007}
{Bregman} J.~N., {Lloyd-Davies} E.~J., 2007, \apj, 669, 990

\bibitem[{{Brunner} {et~al}\mbox{.}(2022){Brunner}, {Liu}, {Lamer}, {Georgakakis}, {Merloni}, {Brusa}, {Bulbul}, {Dennerl}, {Friedrich}, {Liu}, {Maitra}, {Nandra}, {Ramos-Ceja}, {Sanders}, {Stewart}, {Boller}, {Buchner}, {Clerc}, {Comparat}, {Dwelly}, {Eckert}, {Finoguenov}, {Freyberg}, {Ghirardini}, {Gueguen}, {Haberl}, {Kreykenbohm}, {Krumpe}, {Osterhage}, {Pacaud}, {Predehl}, {Reiprich}, {Robrade}, {Salvato}, {Santangelo}, {Schrabback}, {Schwope}, \& {Wilms}}]{Brunner2022}
{Brunner} H. {et~al.}, 2022, \aap, 661, A1

\bibitem[{{Burchett} {et~al}\mbox{.}(2019){Burchett}, {Tripp}, {Prochaska}, {Werk}, {Tumlinson}, {Howk}, {Willmer}, {Lehner}, {Meiring}, {Bowen}, {Bordoloi}, {Peeples}, {Jenkins}, {O'Meara}, {Tejos}, \& {Katz}}]{Burchett2019}
{Burchett} J.~N. {et~al.}, 2019, \apjl, 877, L20

\bibitem[{{Butler Contreras} {et~al}\mbox{.}(2023){Butler Contreras}, {Lau}, {Oppenheimer}, {Bogd{\'a}n}, {Tillman}, {Nagai}, {Kov{\'a}cs}, \& {Burkhart}}]{Contreras2023}
{Butler Contreras} A., {Lau} E.~T., {Oppenheimer} B.~D., {Bogd{\'a}n} {\'A}., {Tillman} M., {Nagai} D., {Kov{\'a}cs} O.~E., {Burkhart} B., 2023, \mnras, 519, 2251

\bibitem[{{Chadayammuri} {et~al}\mbox{.}(2022){Chadayammuri}, {Bogd{\'a}n}, {Oppenheimer}, {Kraft}, {Forman}, \& {Jones}}]{Chadayammuri2022}
{Chadayammuri} U., {Bogd{\'a}n} {\'A}., {Oppenheimer} B.~D., {Kraft} R.~P., {Forman} W.~R., {Jones} C., 2022, \apjl, 936, L15

\bibitem[{{Chawla} {et~al}\mbox{.}(2022){Chawla}, {Kaspi}, {Ransom}, {Bhardwaj}, {Boyle}, {Breitman}, {Cassanelli}, {Cubranic}, {Dong}, {Fonseca}, {Gaensler}, {Giri}, {Josephy}, {Kaczmarek}, {Leung}, {Masui}, {Mena-Parra}, {Merryfield}, {Michilli}, {M{\"u}nchmeyer}, {Ng}, {Patel}, {Pearlman}, {Petroff}, {Pleunis}, {Rahman}, {Sanghavi}, {Shin}, {Smith}, {Stairs}, \& {Tendulkar}}]{Chawla2022}
{Chawla} P. {et~al.}, 2022, \apj, 927, 35

\bibitem[{{Chen} {et~al}\mbox{.}(2020){Chen}, {Zahedy}, {Boettcher}, {Cooper}, {Johnson}, {Rudie}, {Chen}, {Walth}, {Cantalupo}, {Cooksey}, {Faucher-Gigu{\`e}re}, {Greene}, {Lopez}, {Mulchaey}, {Penton}, {Petitjean}, {Putman}, {Rafelski}, {Rauch}, {Schaye}, {Simcoe}, \& {Weiner}}]{Chen20}
{Chen} H.-W. {et~al.}, 2020, \mnras, 497, 498

\bibitem[{{CHIME/FRB Collaboration} {et~al}\mbox{.}(2018){CHIME/FRB Collaboration}, {Amiri}, {Bandura}, {Berger}, {Bhardwaj}, {Boyce}, {Boyle}, {Brar}, {Burhanpurkar}, {Chawla}, {Chowdhury}, {Cliche}, {Cranmer}, {Cubranic}, {Deng}, {Denman}, {Dobbs}, {Fandino}, {Fonseca}, {Gaensler}, {Giri}, {Gilbert}, {Good}, {Guliani}, {Halpern}, {Hinshaw}, {H{\"o}fer}, {Josephy}, {Kaspi}, {Landecker}, {Lang}, {Liao}, {Masui}, {Mena-Parra}, {Naidu}, {Newburgh}, {Ng}, {Patel}, {Pen}, {Pinsonneault-Marotte}, {Pleunis}, {Rafiei Ravandi}, {Ransom}, {Renard}, {Scholz}, {Sigurdson}, {Siegel}, {Smith}, {Stairs}, {Tendulkar}, {Vanderlinde}, \& {Wiebe}}]{Amiri2018}
{CHIME/FRB Collaboration} {et~al.}, 2018, \apj, 863, 48

\bibitem[{{Choudhury}, {Sharma} \& {Quataert}(2019){Choudhury}, {Sharma}, \& {Quataert}}]{Choudhury2019}
{Choudhury} P.~P., {Sharma} P., {Quataert} E., 2019, \mnras, 488, 3195

\bibitem[{{Comparat} {et~al}\mbox{.}(2022){Comparat}, {Truong}, {Merloni}, {Pillepich}, {Ponti}, {Driver}, {Bellstedt}, {Liske}, {Aird}, {Br{\"u}ggen}, {Bulbul}, {Davies}, {Villalba}, {Georgakakis}, {Haberl}, {Liu}, {Maitra}, {Nandra}, {Popesso}, {Predehl}, {Robotham}, {Salvato}, {Thorne}, \& {Zhang}}]{Comparat2022}
{Comparat} J. {et~al.}, 2022, \aap, 666, A156

\bibitem[{{Connor} \& {Ravi}(2022)}]{Connor2022}
{Connor} L., {Ravi} V., 2022, Nature Astronomy

\bibitem[{{Cook} {et~al}\mbox{.}(2023){Cook}, {Bhardwaj}, {Gaensler}, {Scholz}, {Eadie}, {Hill}, {Kaspi}, {Masui}, {Curtin}, {Dong}, {Fonseca}, {Herrera-Martin}, {Kaczmarek}, {Lanman}, {Lazda}, {Leung}, {Meyers}, {Michilli}, {Pandhi}, {Pearlman}, {Pleunis}, {Ransom}, {Rahman}, {Sand}, {Shin}, {Smith}, {Stairs}, \& {Stenning}}]{Cook2023}
{Cook} A.~M. {et~al.}, 2023, arXiv e-prints, arXiv:2301.03502

\bibitem[{{Das} {et~al}\mbox{.}(2021){Das}, {Mathur}, {Gupta}, \& {Krongold}}]{Das2021}
{Das} S., {Mathur} S., {Gupta} A., {Krongold} Y., 2021, \apj, 918, 83

\bibitem[{{Das} {et~al}\mbox{.}(2019){Das}, {Mathur}, {Gupta}, {Nicastro}, {Krongold}, \& {Null}}]{Das2019}
{Das} S., {Mathur} S., {Gupta} A., {Nicastro} F., {Krongold} Y., {Null} C., 2019, \apj, 885, 108

\bibitem[{{Faerman} {et~al}\mbox{.}(2022){Faerman}, {Pandya}, {Somerville}, \& {Sternberg}}]{Faerman2022}
{Faerman} Y., {Pandya} V., {Somerville} R.~S., {Sternberg} A., 2022, \apj, 928, 37

\bibitem[{{Faerman}, {Sternberg} \& {McKee}(2017){Faerman}, {Sternberg}, \& {McKee}}]{Faerman2017}
{Faerman} Y., {Sternberg} A., {McKee} C.~F., 2017, \apj, 835, 52

\bibitem[{{Faerman}, {Sternberg} \& {McKee}(2020){Faerman}, {Sternberg}, \& {McKee}}]{faerman2020}
{Faerman} Y., {Sternberg} A., {McKee} C.~F., 2020, \apj, 893, 82

\bibitem[{{Faerman} \& {Werk}(2023)}]{FW2023}
{Faerman} Y., {Werk} J.~K., 2023, \apj, 956, 92

\bibitem[{{Fang} {et~al}\mbox{.}(2015){Fang}, {Buote}, {Bullock}, \& {Ma}}]{Fang2015}
{Fang} T., {Buote} D., {Bullock} J., {Ma} R., 2015, \apjs, 217, 21

\bibitem[{{Ferland} {et~al}\mbox{.}(2017){Ferland}, {Chatzikos}, {Guzm{\'a}n}, {Lykins}, {van Hoof}, {Williams}, {Abel}, {Badnell}, {Keenan}, {Porter}, \& {Stancil}}]{Ferland2017}
{Ferland} G.~J. {et~al.}, 2017, \rmxaa, 53, 385

\bibitem[{{Flender}, {Nagai} \& {McDonald}(2017){Flender}, {Nagai}, \& {McDonald}}]{Flender2017}
{Flender} S., {Nagai} D., {McDonald} M., 2017, \apj, 837, 124

\bibitem[{{Gnat} \& {Sternberg}(2004)}]{Gnat2004}
{Gnat} O., {Sternberg} A., 2004, \apj, 608, 229

\bibitem[{{Gupta} {et~al}\mbox{.}(2012){Gupta}, {Mathur}, {Krongold}, {Nicastro}, \& {Galeazzi}}]{Gupta2012}
{Gupta} A., {Mathur} S., {Krongold} Y., {Nicastro} F., {Galeazzi} M., 2012, \apjl, 756, L8

\bibitem[{{Haardt} \& {Madau}(2012)}]{Haardt2012}
{Haardt} F., {Madau} P., 2012, \apj, 746, 125

\bibitem[{{Hafen} {et~al}\mbox{.}(2019){Hafen}, {Faucher-Gigu{\`e}re}, {Angl{\'e}s-Alc{\'a}zar}, {Stern}, {Kere{\v{s}}}, {Hummels}, {Esmerian}, {Garrison-Kimmel}, {El-Badry}, {Wetzel}, {Chan}, {Hopkins}, \& {Murray}}]{Hafen2019}
{Hafen} Z. {et~al.}, 2019, \mnras, 488, 1248

\bibitem[{{Henderson} {et~al}\mbox{.}(2016){Henderson}, {Allison}, {Austermann}, {Baildon}, {Battaglia}, {Beall}, {Becker}, {De Bernardis}, {Bond}, {Calabrese}, {Choi}, {Coughlin}, {Crowley}, {Datta}, {Devlin}, {Duff}, {Dunkley}, {D{\"u}nner}, {van Engelen}, {Gallardo}, {Grace}, {Hasselfield}, {Hills}, {Hilton}, {Hincks}, {Hloẑek}, {Ho}, {Hubmayr}, {Huffenberger}, {Hughes}, {Irwin}, {Koopman}, {Kosowsky}, {Li}, {McMahon}, {Munson}, {Nati}, {Newburgh}, {Niemack}, {Niraula}, {Page}, {Pappas}, {Salatino}, {Schillaci}, {Schmitt}, {Sehgal}, {Sherwin}, {Sievers}, {Simon}, {Spergel}, {Staggs}, {Stevens}, {Thornton}, {Van Lanen}, {Vavagiakis}, {Ward}, \& {Wollack}}]{Henderson2016}
{Henderson} S.~W. {et~al.}, 2016, Journal of Low Temperature Physics, 184, 772

\bibitem[{{Hummels} {et~al}\mbox{.}(2019){Hummels}, {Smith}, {Hopkins}, {O'Shea}, {Silvia}, {Werk}, {Lehner}, {Wise}, {Collins}, \& {Butsky}}]{Hummels2019}
{Hummels} C.~B. {et~al.}, 2019, \apj, 882, 156

\bibitem[{{Jankowski} {et~al}\mbox{.}(2023){Jankowski}, {Bezuidenhout}, {Caleb}, {Driessen}, {Malenta}, {Morello}, {Rajwade}, {Sanidas}, {Stappers}, {Surnis}, {Barr}, {Chen}, {Kramer}, {Wu}, {Buchner}, {Serylak}, \& {Prochaska}}]{Jankowski2023}
{Jankowski} F. {et~al.}, 2023, arXiv e-prints, arXiv:2302.10107

\bibitem[{{Johnson}, {Chen} \& {Mulchaey}(2015){Johnson}, {Chen}, \& {Mulchaey}}]{Johnson2015}
{Johnson} S.~D., {Chen} H.-W., {Mulchaey} J.~S., 2015, \mnras, 449, 3263

\bibitem[{{Kraft} {et~al}\mbox{.}(2022){Kraft}, {Markevitch}, {Kilbourne}, {Adams}, {Akamatsu}, {Ayromlou}, {Bandler}, {Bennett}, {Bhardwaj}, {Biffi}, {Bodewits}, {Bogdan}, {Bonamente}, {Borgani}, {Branduardi-Raymont}, {Bregman}, {Burchett}, {Cann}, {Carter}, {Chakraborty}, {Churazov}, {Crain}, {Cumbee}, {Dave}, {DiPirro}, {Dolag}, {Bertrand Doriese}, {Drake}, {Dunn}, {Eckart}, {Eckert}, {Ettori}, {Forman}, {Galeazzi}, {Gall}, {Gatuzz}, {Hell}, {Hodges-Kluck}, {Jackman}, {Jahromi}, {Jennings}, {Jones}, {Kaaret}, {Kavanagh}, {Kelley}, {Khabibullin}, {Kim}, {Koutroumpa}, {Kovacs}, {Kuntz}, {Lin}, {Lau}, {Lee}, {Leutenegger}, {Lisse}, {Lovisari}, {McCammon}, {McEntee}, {Mernier}, {Miller}, {Nagai}, {Negro}, {Nelson}, {Ness}, {Nulsen}, {Ogorzalek}, {Oppenheimer}, {Oskinova}, {Patnaude}, {Pfeifle}, {Pillepich}, {Plucinsky}, {Pooley}, {Porter}, {Randall}, {Rasia}, {Raymond}, {Ruszkowski}, {Sakai}, {Sarkar}, {Sasaki}, {Sato}, {Schellenberger}, {Schaye}, {Simionescu}, {Smith}, {Steiner}, {Stern}, {Su}, {Sun},
  {Tremblay}, {Truong}, {Tutt}, {Veilleux}, {Vikhlinin}, {Vladutescu-Zopp}, {Vogelsberger}, {Walker}, {Weaver}, {Weigt}, {Werk}, {Werner}, {Wolk}, {Zhang}, {Zhang}, {Zhuravleva}, \& {ZuHone}}]{LEM2022}
{Kraft} R. {et~al.}, 2022, arXiv e-prints, arXiv:2211.09827

\bibitem[{{Kravtsov}, {Vikhlinin} \& {Meshcheryakov}(2018){Kravtsov}, {Vikhlinin}, \& {Meshcheryakov}}]{Kravtsov2018}
{Kravtsov} A.~V., {Vikhlinin} A.~A., {Meshcheryakov} A.~V., 2018, Astronomy Letters, 44, 8

\bibitem[{{Lee} {et~al}\mbox{.}(2022){Lee}, {Anbajagane}, {Singh}, {Chluba}, {Nagai}, {Kay}, {Cui}, {Dolag}, \& {Yepes}}]{Lee2022}
{Lee} E. {et~al.}, 2022, \mnras, 517, 5303

\bibitem[{{Lee} {et~al}\mbox{.}(2024){Lee}, {Genel}, {Wandelt}, {Zhang}, {Delgado}, {Pandey}, {Lau}, {Carr}, {Cook}, {Nagai}, {Angles-Alcazar}, {Villaescusa-Navarro}, \& {Bryan}}]{Lee2024}
{Lee} M.~E. {et~al.}, 2024, arXiv e-prints, arXiv:2403.10609

\bibitem[{{Lehner}, {Howk} \& {Wakker}(2015){Lehner}, {Howk}, \& {Wakker}}]{Lehner2015}
{Lehner} N., {Howk} J.~C., {Wakker} B.~P., 2015, \apj, 804, 79

\bibitem[{{Li} {et~al}\mbox{.}(2018){Li}, {Bregman}, {Wang}, {Crain}, \& {Anderson}}]{Li2018}
{Li} J.-T., {Bregman} J.~N., {Wang} Q.~D., {Crain} R.~A., {Anderson} M.~E., 2018, \apjl, 855, L24

\bibitem[{{Lim} {et~al}\mbox{.}(2021){Lim}, {Barnes}, {Vogelsberger}, {Mo}, {Nelson}, {Pillepich}, {Dolag}, \& {Marinacci}}]{Lim2021}
{Lim} S.~H., {Barnes} D., {Vogelsberger} M., {Mo} H.~J., {Nelson} D., {Pillepich} A., {Dolag} K., {Marinacci} F., 2021, \mnras, 504, 5131

\bibitem[{{McQuinn}(2014)}]{McQuinn2014}
{McQuinn} M., 2014, \apjl, 780, L33

\bibitem[{{McQuinn} \& {Werk}(2018)}]{McQuinn2018}
{McQuinn} M., {Werk} J.~K., 2018, \apj, 852, 33

\bibitem[{{Medlock} {et~al}\mbox{.}(2024){Medlock}, {Nagai}, {Singh}, {Oppenheimer}, {Angl{\'e}s Alc{\'a}zar}, \& {Villaescusa-Navarro}}]{Medlock2024}
{Medlock} I., {Nagai} D., {Singh} P., {Oppenheimer} B., {Angl{\'e}s Alc{\'a}zar} D., {Villaescusa-Navarro} F., 2024, arXiv e-prints, arXiv:2403.02313

\bibitem[{{Moser} {et~al}\mbox{.}(2022){Moser}, {Battaglia}, {Nagai}, {Lau}, {Machado Poletti Valle}, {Villaescusa-Navarro}, {Amodeo}, {Angl{\'e}s-Alc{\'a}zar}, {Bryan}, {Dave}, {Hernquist}, \& {Vogelsberger}}]{Moser2022}
{Moser} E. {et~al.}, 2022, \apj, 933, 133

\bibitem[{{Mowla} {et~al}\mbox{.}(2019){Mowla}, {van der Wel}, {van Dokkum}, \& {Miller}}]{Mowla2019}
{Mowla} L., {van der Wel} A., {van Dokkum} P., {Miller} T.~B., 2019, \apjl, 872, L13

\bibitem[{{Nelson}, {Lau} \& {Nagai}(2014){Nelson}, {Lau}, \& {Nagai}}]{Nelson2014}
{Nelson} K., {Lau} E.~T., {Nagai} D., 2014, \apj, 792, 25

\bibitem[{{Newburgh} {et~al}\mbox{.}(2016){Newburgh}, {Bandura}, {Bucher}, {Chang}, {Chiang}, {Cliche}, {Dav{\'e}}, {Dobbs}, {Clarkson}, {Ganga}, {Gogo}, {Gumba}, {Gupta}, {Hilton}, {Johnstone}, {Karastergiou}, {Kunz}, {Lokhorst}, {Maartens}, {Macpherson}, {Mdlalose}, {Moodley}, {Ngwenya}, {Parra}, {Peterson}, {Recnik}, {Saliwanchik}, {Santos}, {Sievers}, {Smirnov}, {Stronkhorst}, {Taylor}, {Vanderlinde}, {Van Vuuren}, {Weltman}, \& {Witzemann}}]{Newburgh2016}
{Newburgh} L.~B. {et~al.}, 2016, in Society of Photo-Optical Instrumentation Engineers (SPIE) Conference Series, Vol. 9906, Ground-based and Airborne Telescopes VI, {Hall} H.~J., {Gilmozzi} R., {Marshall} H.~K., eds., p. 99065X

\bibitem[{{Ni} {et~al}\mbox{.}(2023){Ni}, {Genel}, {Angl{\'e}s-Alc{\'a}zar}, {Villaescusa-Navarro}, {Jo}, {Bird}, {Di Matteo}, {Croft}, {Chen}, {de Santi}, {Gebhardt}, {Shao}, {Pandey}, {Hernquist}, \& {Dave}}]{Ni2023}
{Ni} Y. {et~al.}, 2023, \apj, 959, 136

\bibitem[{{Oppenheimer}(2018)}]{Oppenheimer2018}
{Oppenheimer} B.~D., 2018, \mnras, 480, 2963

\bibitem[{{Osato} \& {Nagai}(2022)}]{osato22}
{Osato} K., {Nagai} D., 2022, arXiv e-prints, arXiv:2201.02632

\bibitem[{{Ostriker}, {Bode} \& {Babul}(2005){Ostriker}, {Bode}, \& {Babul}}]{ostriker05}
{Ostriker} J.~P., {Bode} P., {Babul} A., 2005, \apj, 634, 964

\bibitem[{{Pandya} {et~al}\mbox{.}(2023){Pandya}, {Fielding}, {Bryan}, {Carr}, {Somerville}, {Stern}, {Faucher-Gigu{\`e}re}, {Hafen}, {Angl{\'e}s-Alc{\'a}zar}, \& {Forbes}}]{Pandya2022}
{Pandya} V. {et~al.}, 2023, \apj, 956, 118

\bibitem[{{Peeples} {et~al}\mbox{.}(2019){Peeples}, {Corlies}, {Tumlinson}, {O'Shea}, {Lehner}, {O'Meara}, {Howk}, {Earl}, {Smith}, {Wise}, \& {Hummels}}]{Peeples19}
{Peeples} M.~S. {et~al.}, 2019, \apj, 873, 129

\bibitem[{{Prochaska} \& {Zheng}(2019)}]{PZ19}
{Prochaska} J.~X., {Zheng} Y., 2019, \mnras, 485, 648

\bibitem[{{Qu} \& {Bregman}(2018)}]{Qu2018}
{Qu} Z., {Bregman} J.~N., 2018, \apj, 856, 5

\bibitem[{{Qu} {et~al}\mbox{.}(2024){Qu}, {Chen}, {Johnson}, {Rudie}, {Zahedy}, {DePalma}, {Schaye}, {Boettcher}, {Cantalupo}, {Chen}, {Faucher-Gigu{\`e}re}, {Li}, {Mulchaey}, {Petitjean}, \& {Rafelski}}]{Qu2024}
{Qu} Z. {et~al.}, 2024, arXiv e-prints, arXiv:2402.08016

\bibitem[{{Ramesh} \& {Nelson}(2024)}]{Ramesh2023}
{Ramesh} R., {Nelson} D., 2024, \mnras, 528, 3320

\bibitem[{{Ravi}(2019)}]{Ravi2019}
{Ravi} V., 2019, \apj, 872, 88

\bibitem[{{Ravi} {et~al}\mbox{.}(2023){Ravi}, {Catha}, {Chen}, {Connor}, {Cordes}, {Faber}, {Lamb}, {Hallinan}, {Harnach}, {Hellbourg}, {Hobbs}, {Hodge}, {Hodges}, {Law}, {Rasmussen}, {Sharma}, {Sherman}, {Shi}, {Simard}, {Somalwar}, {Squillace}, {Weinreb}, {Woody}, \& {Yadlapalli}}]{Ravi2023}
{Ravi} V. {et~al.}, 2023, arXiv e-prints, arXiv:2301.01000

\bibitem[{{Ried Guachalla} {et~al}\mbox{.}(2023){Ried Guachalla}, {Schaan}, {Hadzhiyska}, \& {Ferraro}}]{Ried2023}
{Ried Guachalla} B., {Schaan} E., {Hadzhiyska} B., {Ferraro} S., 2023, arXiv e-prints, arXiv:2312.12435

\bibitem[{{Schaan} {et~al}\mbox{.}(2021){Schaan}, {Ferraro}, {Amodeo}, {Battaglia}, {Aiola}, {Austermann}, {Beall}, {Bean}, {Becker}, {Bond}, {Calabrese}, {Calafut}, {Choi}, {Denison}, {Devlin}, {Duff}, {Duivenvoorden}, {Dunkley}, {D{\"u}nner}, {Gallardo}, {Guan}, {Han}, {Hill}, {Hilton}, {Hilton}, {Hlo{\v{z}}ek}, {Hubmayr}, {Huffenberger}, {Hughes}, {Koopman}, {MacInnis}, {McMahon}, {Madhavacheril}, {Moodley}, {Mroczkowski}, {Naess}, {Nati}, {Newburgh}, {Niemack}, {Page}, {Partridge}, {Salatino}, {Sehgal}, {Schillaci}, {Sif{\'o}n}, {Smith}, {Spergel}, {Staggs}, {Storer}, {Trac}, {Ullom}, {Van Lanen}, {Vale}, {van Engelen}, {Maga{\~n}a}, {Vavagiakis}, {Wollack}, {Xu}, \& {Atacama Cosmology Telescope Collaboration}}]{Schaan2021}
{Schaan} E. {et~al.}, 2021, \prd, 103, 063513

\bibitem[{{Scott} {et~al}\mbox{.}(2023){Scott}, {Cho}, {Day}, {Deller}, {Glowacki}, {Gourdji}, {Bannister}, {Bera}, {Bhandari}, {James}, \& {Shannon}}]{Scott2023}
{Scott} D.~R. {et~al.}, 2023, arXiv e-prints, arXiv:2301.13484

\bibitem[{{Shaw} {et~al}\mbox{.}(2010){Shaw}, {Nagai}, {Bhattacharya}, \& {Lau}}]{Shaw2010}
{Shaw} L.~D., {Nagai} D., {Bhattacharya} S., {Lau} E.~T., 2010, \apj, 725, 1452

\bibitem[{{Shi}(2016)}]{Shi2016}
{Shi} X., 2016, \mnras, 461, 1804

\bibitem[{{Shirasaki}, {Lau} \& {Nagai}(2020){Shirasaki}, {Lau}, \& {Nagai}}]{shirasaki_etal20}
{Shirasaki} M., {Lau} E.~T., {Nagai} D., 2020, \mnras, 491, 235

\bibitem[{{Singh}, {Voit} \& {Nath}(2021){Singh}, {Voit}, \& {Nath}}]{singh2021}
{Singh} P., {Voit} G.~M., {Nath} B.~B., 2021, \mnras, 501, 2467

\bibitem[{{Smith} {et~al}\mbox{.}(2016){Smith}, {Abraham}, {Allured}, {Bautz}, {Bookbinder}, {Bregman}, {Brenneman}, {Brickhouse}, {Burrows}, {Burwitz}, {Carvalho}, {Cheimets}, {Costantini}, {Dawson}, {DeRoo}, {Falcone}, {Foster}, {Grant}, {Heilmann}, {Hertz}, {Hine}, {Huenemoerder}, {Kaastra}, {Madsen}, {McEntaffer}, {Miller}, {Miller}, {Morse}, {Mushotzky}, {Nandra}, {Nowak}, {Paerels}, {Petre}, {Plice}, {Poppenhaeger}, {Ptak}, {Reid}, {Sanders}, {Schattenburg}, {Schulz}, {Smale}, {Temi}, {Valencic}, {Walker}, {Willingale}, {Wilms}, \& {Wolk}}]{Smith2016}
{Smith} R.~K. {et~al.}, 2016, in Society of Photo-Optical Instrumentation Engineers (SPIE) Conference Series, Vol. 9905, Space Telescopes and Instrumentation 2016: Ultraviolet to Gamma Ray, {den Herder} J.-W.~A., {Takahashi} T., {Bautz} M., eds., p. 99054M

\bibitem[{{Stern} {et~al}\mbox{.}(2018){Stern}, {Faucher-Gigu{\`e}re}, {Hennawi}, {Hafen}, {Johnson}, \& {Fielding}}]{Stern2018}
{Stern} J., {Faucher-Gigu{\`e}re} C.-A., {Hennawi} J.~F., {Hafen} Z., {Johnson} S.~D., {Fielding} D., 2018, \apj, 865, 91

\bibitem[{{Stern} {et~al}\mbox{.}(2019){Stern}, {Fielding}, {Faucher-Gigu{\`e}re}, \& {Quataert}}]{Stern2019}
{Stern} J., {Fielding} D., {Faucher-Gigu{\`e}re} C.-A., {Quataert} E., 2019, \mnras, 488, 2549

\bibitem[{{Stern} {et~al}\mbox{.}(2020){Stern}, {Fielding}, {Faucher-Gigu{\`e}re}, \& {Quataert}}]{Stern2020}
{Stern} J., {Fielding} D., {Faucher-Gigu{\`e}re} C.-A., {Quataert} E., 2020, \mnras, 492, 6042

\bibitem[{{Stern} {et~al}\mbox{.}(2023){Stern}, {Fielding}, {Hafen}, {Su}, {Naor}, {Faucher-Gigu{\`e}re}, {Quataert}, \& {Bullock}}]{Stern2023}
{Stern} J., {Fielding} D., {Hafen} Z., {Su} K.-Y., {Naor} N., {Faucher-Gigu{\`e}re} C.-A., {Quataert} E., {Bullock} J., 2023, arXiv e-prints, arXiv:2306.00092

\bibitem[{{Sunyaev} \& {Zeldovich}(1972)}]{SZ1972}
{Sunyaev} R.~A., {Zeldovich} Y.~B., 1972, Comments on Astrophysics and Space Physics, 4, 173

\bibitem[{{Tanimura} {et~al}\mbox{.}(2022){Tanimura}, {Aghanim}, {Bonjean}, \& {Zaroubi}}]{Tanimura22}
{Tanimura} H., {Aghanim} N., {Bonjean} V., {Zaroubi} S., 2022, \aap, 662, A48

\bibitem[{{Tchernyshyov} {et~al}\mbox{.}(2022){Tchernyshyov}, {Werk}, {Wilde}, {Prochaska}, {Tripp}, {Burchett}, {Bordoloi}, {Howk}, {Lehner}, {O'Meara}, {Tejos}, \& {Tumlinson}}]{Tchernyshyov2022}
{Tchernyshyov} K. {et~al.}, 2022, \apj, 927, 147

\bibitem[{{Tumlinson}, {Peeples} \& {Werk}(2017){Tumlinson}, {Peeples}, \& {Werk}}]{Tumlinson2017}
{Tumlinson} J., {Peeples} M.~S., {Werk} J.~K., 2017, \araa, 55, 389

\bibitem[{{Tumlinson} {et~al}\mbox{.}(2011){Tumlinson}, {Thom}, {Werk}, {Prochaska}, {Tripp}, {Weinberg}, {Peeples}, {O'Meara}, {Oppenheimer}, {Meiring}, {Katz}, {Dav{\'e}}, {Ford}, \& {Sembach}}]{Tumlinson2011}
{Tumlinson} J. {et~al.}, 2011, Science, 334, 948

\bibitem[{{van de Voort} {et~al}\mbox{.}(2019){van de Voort}, {Springel}, {Mandelker}, {van den Bosch}, \& {Pakmor}}]{Voort19}
{van de Voort} F., {Springel} V., {Mandelker} N., {van den Bosch} F.~C., {Pakmor} R., 2019, \mnras, 482, L85

\bibitem[{{Villaescusa-Navarro} {et~al}\mbox{.}(2021){Villaescusa-Navarro}, {Angl{\'e}s-Alc{\'a}zar}, {Genel}, {Spergel}, {Somerville}, {Dave}, {Pillepich}, {Hernquist}, {Nelson}, {Torrey}, {Narayanan}, {Li}, {Philcox}, {La Torre}, {Maria Delgado}, {Ho}, {Hassan}, {Burkhart}, {Wadekar}, {Battaglia}, {Contardo}, \& {Bryan}}]{Villaescusa2021}
{Villaescusa-Navarro} F. {et~al.}, 2021, \apj, 915, 71

\bibitem[{{Voit}(2005)}]{Voit2005}
{Voit} G.~M., 2005, Reviews of Modern Physics, 77, 207

\bibitem[{{Voit}(2019)}]{voit2019_column_densities}
{Voit} G.~M., 2019, \apj, 880, 139

\bibitem[{{Voit} {et~al}\mbox{.}(2018){Voit}, {Ma}, {Greene}, {Goulding}, {Pandya}, {Donahue}, \& {Sun}}]{voit2018_precipitation}
{Voit} G.~M., {Ma} C.~P., {Greene} J., {Goulding} A., {Pandya} V., {Donahue} M., {Sun} M., 2018, \apj, 853, 78

\bibitem[{{Voit} {et~al}\mbox{.}(2017){Voit}, {Meece}, {Li}, {O'Shea}, {Bryan}, \& {Donahue}}]{Voit2017}
{Voit} G.~M., {Meece} G., {Li} Y., {O'Shea} B.~W., {Bryan} G.~L., {Donahue} M., 2017, \apj, 845, 80

\bibitem[{{Werk} {et~al}\mbox{.}(2013){Werk}, {Prochaska}, {Thom}, {Tumlinson}, {Tripp}, {O'Meara}, \& {Peeples}}]{Werk2014}
{Werk} J.~K., {Prochaska} J.~X., {Thom} C., {Tumlinson} J., {Tripp} T.~M., {O'Meara} J.~M., {Peeples} M.~S., 2013, \apjs, 204, 17

\bibitem[{{Wijers}, {Schaye} \& {Oppenheimer}(2020){Wijers}, {Schaye}, \& {Oppenheimer}}]{Wijers2020}
{Wijers} N.~A., {Schaye} J., {Oppenheimer} B.~D., 2020, \mnras, 498, 574

\bibitem[{{Wu} \& {McQuinn}(2022)}]{Wu2022}
{Wu} X., {McQuinn} M., 2022, arXiv e-prints, arXiv:2209.04455

\bibitem[{{Yang} {et~al}\mbox{.}(2022){Yang}, {Cai}, {Cui}, {Dav{\'e}}, {Peacock}, \& {Sorini}}]{Yang2022}
{Yang} T., {Cai} Y.-C., {Cui} W., {Dav{\'e}} R., {Peacock} J.~A., {Sorini} D., 2022, \mnras, 516, 4084

\bibitem[{{Zhang} {et~al}\mbox{.}(2024){Zhang}, {Comparat}, {Ponti}, {Meloni}, {Nandra}, {Haberl}, {Locatelli}, {Zhang}, {Sanders}, {Zheng}, {Liu}, {Popesso}, {Liu}, {Truong}, {Pillepich}, {Predehl}, \& {Salvato}}]{Zhang2024}
{Zhang} Y. {et~al.}, 2024, arXiv e-prints, arXiv:2401.17308

\end{thebibliography}
		\bibliographystyle{mn2e}
	}

\begin{appendices}
\renewcommand\thefigure{\thesection.\arabic{figure}}    
\renewcommand{\thetable}{A\arabic{table}}
\renewcommand{\theequation}{A\arabic{equation}}

\section{Observable dependence on model-specific parameters}
\setcounter{figure}{0}    
\setcounter{table}{0}
\setcounter{equation}{0}
\label{apn-model-unc}
In Section~\ref{sec-multi-wavelength}, we compared the predictions of the CGM models using the fiducial values of model parameters. Many of these model parameters significantly affect the normalisation and shape of the observed profiles. In Figure~\ref{fig-Obs-profiles-unc}, we include a range in the model-specific parameters to highlight their impact on observable quantities. We list the parameters varied and their ranges in Table~\ref{tab-parameter-depn}\footnote{For the isentropic model, we only show the variation in $\alpha$ in Figure~\ref{fig-Obs-profiles-unc} for simplicity.}. 

In addition, we compile many of the currently available CGM observations and show them in Figure~\ref{fig-Obs-profiles-unc}. Note that some of these measurements are limited to more massive galaxies than our fiducial choice, span a range in redshifts, and may suffer from Malmquist bias due to the stacking procedure, and therefore are meant for a qualitative rather than a quantitative comparison with the model predictions. 

\begin{figure*}
\centering
	\includegraphics[height=17cm,angle=0.0]{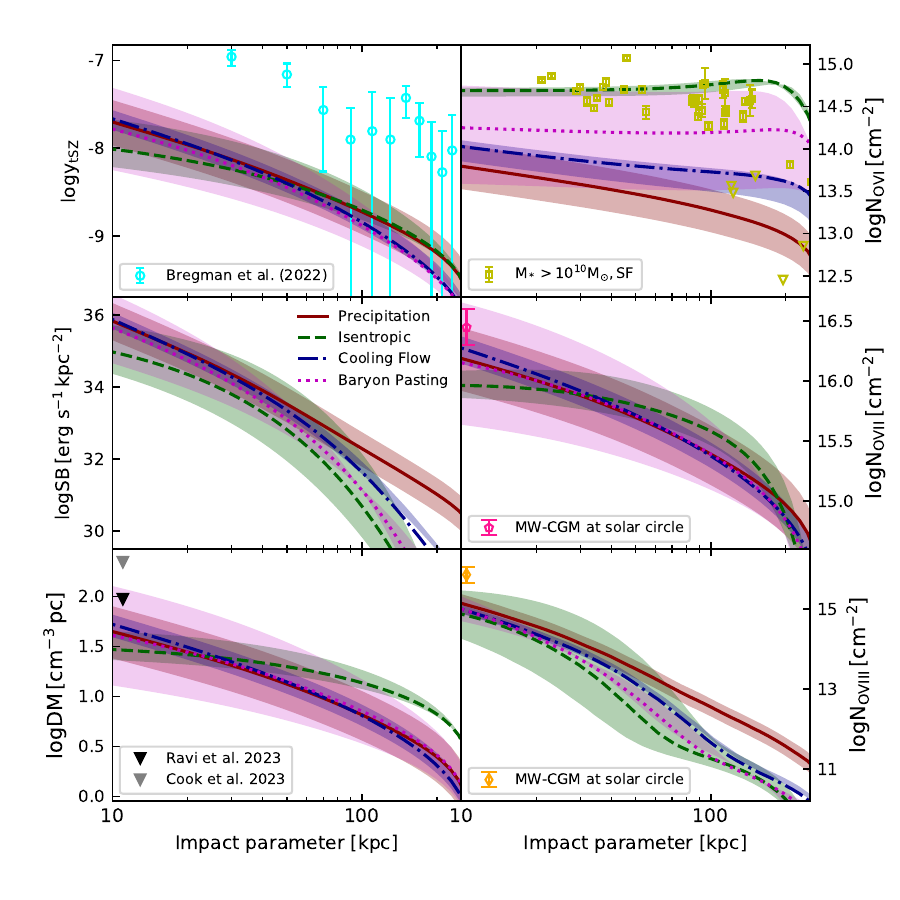}
	\caption{Same as Figure \ref{fig-Obs-profiles} with the shaded regions around each of the predictions of the fiducial model representing the uncertainties in the model-specific parameters as discussed in Section \ref{apn-model-unc}. The cyan circles in the top-left panel show the stacked tSZ signal for $L_*$ galaxies \citep{Bregman2022}. The yellow squares (triangles) in the top-right panel show OVI column density measurements (upper limits) from COS-Halos \citep{Tumlinson2011} and eCGM surveys \citep{Johnson2015}. The pink hexagon (middle-right panel) and orange diamond (bottom-right panel) are the OVII \citep{Bregman2007, Fang2015, Das2019} and OVIII \citep{Gupta2012, Das2019} column density measurements, respectively, for MW at the solar circle, multiplied by a factor of two to compare with model predictions of projected column densities as an external observer. In the lower left-hand panel, black and grey triangles represent the upper limits of the MW-CGM dispersion measure from a localised FRB in the Deep Synoptic Array \citep{Ravi2023} and the CHIME-FRB catalogue \citep{Cook2023}, respectively, for an external observer at the solar circle.}
	\label{fig-Obs-profiles-unc}
\end{figure*}

The tSZ data (cyan circles in the top-left panel) are taken from \cite{Bregman2022} (see Figure~8). They represent the dimensionless $y$-parameter measured from the {\it Planck} and {\it WMAP} datasets by stacking 11 $L_*$ galaxies in the local Universe with distance $<10$ Mpc (i.e. $z<0.003$). The median stellar mass of the sample is $\sim 6.8\times 10^{10}~M_{\odot}$ corresponding to an approximate virial mass $\sim 2 \times 10^{12}~M_{\odot}$ (two times the fiducial mass assumed for the models in our work). All four fiducial CGM models lie an order of magnitude below the tSZ signal, likely driven by the more massive and hotter halos in the observational dataset ($y \propto M_{\rm vir}^{^{5/3}}$). 

The OVI absorption line column densities (yellow squares and triangles in the top-right panel) are taken from COS-Halos \citep{Tumlinson2011} and eCGM surveys \citep{Johnson2015}, using only absorption features associated with late-type, isolated galaxies with stellar masses above $10^{10}~M_{\odot}$. These galaxies span a redshift range $0.1-0.4$ with a median stellar mass of $2.8\times 10^{10}~M_{\odot}$ corresponding to a median halo mass of $\sim10^{12}~M_{\odot}$ \citep{Kravtsov2018, Mowla2019}. The fiducial isentropic model is calibrated to reproduce $N_{\rm OVI}$, whereas other models are not, lying one to two orders of magnitude below the column density measurements. \cite{Qu2024} report OVI absorption in the CGM of galaxies at $0.4<z<0.7$. For the massive star forming galaxies in their sample, they find that the combined column density profile can be fit with $\log(\rm N_{OVI}(b=R_{200}))=14.20 \pm 0.09$ and a slope of $0.74 \pm 0.21$ (see their Figure 9 and Table 2). The baryon pasting and isentropic models produce similar column densities and flatter profiles, while the precipitation model has a more similar slope and lower column densities. We note that the galaxy sample described by \cite{Qu2024} is at higher redshifts than our models, and the slope they infer might be affected by column densities at impact parameters $\approx r_{200}$.

The column densities of OVII (pink hexagon in the middle-right panel) and OVIII (orange diamond in the bottom-right panel) are the measurements of MW-CGM in the solar circle \citep{Gupta2012, Bregman2007, Fang2015, Das2019}. We multiplied the observed column densities by a factor of two to mimic the projected column densities as an external observer. As shown in Figure~\ref{fig-Obs-profiles-unc}, the precipitation, cooling flow, and baryon pasting models produce higher OVII columns due to their higher central densities than the isentropic model. All four fiducial models underestimate OVIII. We note that the fiducial isentropic model described in \citetalias{faerman2020} has a solar metallicity in the inner CGM, and matches the OVII and OVIII columns measured in the MW (see their Table~2). This demonstrates that the ion column densities are sensitive to modelling assumptions such as ionisation equilibrium, metallicity distribution, and an external radiation field (e.g., see Figure~6 in \citealt{Faerman2022}). A full exploration of how individual model predictions vary with such modelling assumptions is beyond the scope of this work.

\cite{Ravi2023} derived an upper limit on the DM contribution by the MW-CGM using a recently discovered non-repeating FRB (20220319D) from the Deep Synoptic Array. The limit is shown by the black triangle in the lower-left panel of Figure~\ref{fig-Obs-profiles-unc} for an external observer with the impact parameter passing through the solar circle. The upper limit can decrease by 40\% depending on the value they use for ISM contamination of the DM. The grey triangle in the same panel corresponds to the upper limit derived by \cite{Cook2023} using the CHIME-FRB catalogue. Their upper limit can decrease by 50\% depending on their ISM contamination model. All four CGM models considered here are consistent with the two measurements.

We also attempt to capture the dependence of tSZ, DM\footnote{We do not show the parameter dependencies for kSZ since it's identical to DM.}, soft X-ray emission, and oxygen absorption column densities on the model parameters near the inner ($\sim 50$ kpc) and outer ($\sim 150$ kpc) CGM. To quantify the dependence of a given observable $O$ on a particular model parameter $P$, we calculate the power law slope $\alpha_{\rm PL}$ defined as $O \propto P^{\alpha_{\rm PL}}$ (with fixed boundary conditions). We tabulate the values $\alpha_{\rm PL}$ in Table \ref{tab-parameter-depn}.

In the case of the precipitation model, the X-ray emission is more sensitive to the value of $t_{\rm cool}/t_{\rm ff}$ with $\alpha_{\rm PL} \sim -1.6$, while other observables show very similar values of $\alpha_{\rm PL} (\sim -0.8)$, making the X-ray emission the most optimal tool for constraining $t_{\rm cool}/t_{\rm ff}$. These trends are driven by the density dependence of the observables, since the temperature does not vary significantly due to the fixed boundary condition. The sensitivity to $t_{\rm cool}/t_{\rm ff}$ does not depend on the radial distance of any observables considered here. The cooling flow model mimics the trends seen in the precipitation model with $\alpha_{\rm PL} \sim 1$ for X-ray emission and $\sim 0.5$ for other observables. These results further highlight the need for deeper X-ray surveys. In the case of baryon pasting, $\epsilon_f$ has the strongest impact on X-ray emission ($\alpha_{\rm PL} \sim -1.6$ at 50~kpc and $\sim 2.4$ at 150 kpc). For most other observables, the levels of sensitivity range from $\alpha_{\rm PL} \sim -1$ to $-2$ (except OVIII). The isentropic model shows more interesting trends as a function of the input parameters and the radial range. For example, $N_{\rm OVI}$ shows an opposite trend to other observables as a function of $\sigma_{\rm turb}$ and $\alpha$. X-ray emission and $N_{\rm OVIII}$ are most sensitive to the model parameters. These trends are due to a peak in their respective ion fractions in a narrow temperature range (see \citealp{Faerman2022} for more details).

The purpose of the analysis and the results shown in Figure~\ref{fig-Obs-profiles-unc} and Table~\ref{tab-parameter-depn} is to identify the model parameters to which the observable quantities are most sensitive, helping in planning the most effective strategy to select observations to constrain CGM physics.

\begin{table*}
\caption{Dependence of observable quantities $(\alpha_{\rm PL}$) on model-specific input parameters near the inner ($\approx$ 50 kpc) and the outer ($\approx$ 150 kpc) CGM.}
\centering
\resizebox{0.8 \textwidth}{!}{
\setlength{\tabcolsep}{6pt}
\begin{tabular}{ c c c c c c}
& & Precipitation & Isentropic & Cooling Flow & Baryon pasting \\
\toprule 
 &  & $t_{\rm cool}/t_{\rm ff}$ & $\sigma_{\rm turb}$\hspace{5mm} $\alpha$  & $ \rm \Dot{M}$ & $10^6\epsilon_f$ \\
 &  & $[5,20]$ & $[20,100]$\hspace{3mm} $[1,3]$  & $[0.5,1.5]$ & $[0.5,1.5]$ \\
 &  & & $\rm km s^{-1}$\hspace{6mm}  & $\rm M_{\odot}\, y^{-1}$ &  \\
 \midrule
tSZ & 50 kpc & $-0.8$ & $-0.4$ \hspace{5mm} $-0.9$ & $0.5$ &  $-1.2$ \\\\
 & 150 kpc &  & $-0.2$ \hspace{5mm} $-0.3$  &  & $-0.7$ \\
\midrule
DM  & 50 kpc & $-0.8$ & $-0.2$ \hspace{5mm} $-0.5$  & $0.5$ & $-1.4$ \\\\
 & 150 kpc &  & $-0.1$ \hspace{5mm} $-0.2$  &  & $-1$\\
 \midrule
 XSB [0.5-2.0]~keV & 50 kpc & $-1.6$ & $-1.3$ \hspace{5mm} $-3.7$ & $1$ & $-1.6$\\\\
 & 150 kpc &  & $-1.5$ \hspace{5mm} $-4.1$ & & $2.4$ \\
 \midrule
$\rm N_{OVI}$ & 50 kpc & $-0.9$ & $0.1$ \hspace{9mm} $0.3$ & $0.5$ & $-1.7$\\\\
 & 150 kpc &  & $0.1$ \hspace{5mm} $-0.1$ & & $-1.6$ \\
 \midrule
$\rm N_{OVII}$ & 50 kpc & $-0.8$ & $-0.3$\hspace{5mm} $-0.7$ & $0.5$ & $-1.3$ \\\\
 & 150 kpc &  & $-0.4$ \hspace{5mm} $-0.9$ &  &  $-0.7$\\
 \midrule
$\rm N_{OVIII}$ & 50 kpc & $-0.8$ & $-1.4$\hspace{5mm} $-4.3$ & $0.5$ & $1$ \\\\
 & 150 kpc &  & $-0.6$  \hspace{5mm} $-1.1$ &  & $\approx 0$ \\
\bottomrule
\end{tabular}}
\label{tab-parameter-depn}
\end{table*}

\end{appendices}
\end{document}